\documentclass[fleqn,usenatbib]{mnras}

\usepackage{newtxtext,newtxmath}

\usepackage[T1]{fontenc}

\DeclareRobustCommand{\VAN}[3]{#2}
\let\VANthebibliography\thebibliography
\def\thebibliography{\DeclareRobustCommand{\VAN}[3]{##3}\VANthebibliography}

\usepackage{graphicx}	
\usepackage{amsmath}	

\usepackage{amssymb}	
\usepackage{pdflscape}  
\usepackage{subcaption}

\title[UCDs ejected from disintegrating systems]{Identifying and characterizing ultracool dwarfs ejected from post-encounter disintegrating systems}

\author[A. K. P. Yip et al.]{Alexandra K. P. Yip,$^{1,2}$\thanks{E-mail: alexandra.yip@postgrado.uv.cl (AY)}
Radostin Kurtev,$^{1,2}$
David J. Pinfield,$^{3}$
Federico Marocco,$^{4}$ \newauthor
Mariusz Gromadzki,$^{5}$
and Julio A. Carballo-Bello$^{6}$
\\
$^{1}$Departamento de Fisica y Astronomia, Facultad de Ciencias, Universidad de Valparaiso, Av. Gran Bretana 1111, Casilla 5030, Valparaiso, Chile\\
$^{2}$Millennium Institute of Astrophysics, Nuncio Monsenor Sotero Sanz 100, Of. 104, Providencia, Santiago, Chile\\
$^{3}$Centre for Astrophysics Research, Department of Physics, Astronomy and Mathematics, University of Hertfordshire, Hatfield AL10 9AB\\
$^{4}$IPAC, Mail Code 100-22, Caltech, 1200 E. California Blvd., Pasadena, CA 91125, USA\\
$^{5}$Astronomical Observatory, University of Warsaw, Al. Ujazdowskie 4, 00-478 Warszawa, Poland \\
$^{6}$Instituto de Alta Investigaci\'on, Sede Esmeralda, Universidad de Tarapac\'a, Av. Luis Emilio Recabarren 2477, Iquique, Chile
}

\date{Accepted 2023 April 3. Received 2023 April 3; in original form 2023 February 15}

\pubyear{2023}

\begin{document}
\label{firstpage}
\pagerange{\pageref{firstpage}--\pageref{lastpage}}
\maketitle

\begin{abstract}
Disintegrating multiple systems have been previously discovered from kinematic studies of the \textit{Hipparcos} catalogue. They are presumably the result of dynamical encounters taking place in the Galactic disk between single/multiple systems. In this paper, we aim to expand the search for such systems, to study their properties, as well as to characterize possible low-mass ejecta (i.e. brown dwarfs and planets). We have assembled a list of 15 candidate systems using astrometry from the Tycho-Gaia astrometric solution (later upgraded with \textit{Gaia} DR3), and here we present the discovery and follow-up of 5 of them. We have obtained DECam imaging for all 5 systems and by combining near-infrared photometry and proper motion, we searched for ultra-cool ejected components. We find that the system consisting of TYC 7731-1951-1, TYC 7731-2128 AB, and TYC 7731-1995-1ABC?, contains one very promising ultra-cool dwarf candidate. Using additional data from the literature, we have found that 3 out of 5 disintegrating system candidates are likely to be true disintegrating systems.   
\end{abstract}




\section{Introduction}

Disintegrating multiple system can help us fathom the formation and evolution of binaries and multiple systems \citep[e.g.][]{1972AJ.....77..169S}. This in turn can lead to a more complete picture of the mechanism of star formation \citep{2000AJ....120.3177R}. In particular cases, when a binary or multiple system interacts with another close by star or multiple system the binding energy between the components is gradually reduced, causing the system to become unbound. Systems with a large cross-section (i.e. wide binaries and multiple systems) are more likely to interact with stellar or sub-stellar objects. Normally, such interaction are relatively weak. However, there are more violent interactions that can lead to the break-up of the system or all of the systems involved \citep{1975MNRAS.173..729H,1987ApJ...312..367W}.

Disintegrating systems are thought to be common \citep{2009A&A...504..277L}, but the average time for disintegration is short, approximately 1 Myr \citep{1972AJ.....77..169S} therefore disintegrating systems are very difficult to identify. Studies in this area have so far been limited, and have generally focused on kinematic studies \citep[e.g.][]{1982ApJ...254..214R} or numerical simulations \citep[e.g.][]{1983MNRAS.203.1107M,1987ApJ...312..367W}. \citet{1990tbp..book.....M} developed an analytical technique to identify escaped objects from N-body systems. The test involves only a one-dimensional projected motion vector of the system at any given time. This makes it well-suited to study stellar systems of which the full motion vectors are unknown. Later, \citet{2009A&A...504..277L} attempted to adapt the Marchal’s test to real triple stars from the \textit{HIgh Precision PARallax COllecting Satellite} catalogue \citep[\textit{Hipparcos};][]{1997A&A...323L..49P,2007ASSL..350.....V}. 

Triple systems are found at a higher rate in hierarchical configuration (i.e. with two stars forming a tight pair and a third component on a wide orbit) rather than in random arrangements \citep{2007MNRAS.379..111V,2016ComAC...3....6T}. This suggests that systems not in hierarchical configuration are likely unstable or, at least, less stable than hierarchical systems. Nevertheless hierarchical systems can become unstable too. Numerical simulations on systems with hierarchical configurations have shown that $\sim$95\% of them have had an ejection in their life-time \citep[e.g.][]{1990CeMDA..48..357A}. Moreover triple systems can form via binary-binary interactions, leading to the ejection of one of the component of the original binaries \citep{1974ApJ...190..253S}. This is inferred from the dynamical evolution of stellar clusters \citep[e.g.][]{2003IAUS..208..295A,2004RMxAC..21..156A} and is predicted by numerical simulations on binary-binary encounters \citep{1983MNRAS.203.1107M}. Finally, since the hierarchical configuration appears to be the more stable layout, newly formed triple systems rearrange into a hierarchical arrangement shortly after their formation \citep{2009A&A...504..277L}. Therefore, many of the hierarchical triple systems we observe today are likely to reach a disintegration phase shortly or to be undergoing that process currently. The Marchal's test is not universally applicable. For example, in the case of binary-third body systems their period often exceeds the baseline covered by available observations, therefore leading to unreliable orbital solutions \citep{2005ESASP.576..251L}. 

The Marchal's test is a good tool to identify real disintegrating systems because it only requires the one-dimensional projected motion and usually the full motion of stars is not known. \citet{2009A&A...504..277L} designed an algorithm based on the Marchal's test and applied it to 24 \textit{Hipparcos} triple systems finding that 10 out of 24 will have an unavoidable escape event. 

Multiple systems can also disintegrate as a result of the evolution of one of the component from main sequence star to white dwarf \citep{2014MNRAS.437.1127V,2018MNRAS.480.4884E}. Wide binaries and multiple systems are particularly vulnerable \citep{2014MNRAS.437.1127V}, because the binding energy of the system is low and therefore even weaker interactions such as Galactic tides can lead to a break up. 
Studies of disintegrating multiple systems need very precise parallaxes and proper motions (hereafter PM) to be able to distinguish between genuine systems and mere chance alignments. The lack of such measurements has limited the studies of these important systems but now with the advent of \textit{Gaia} \citep{2016A&A...595A...1G,2021A&A...649A...1G} there has never been a better time to look for such extremely rare yet unquestionably interesting systems. Using Gaia DR2, \citet{2018yCat..36160037B} searched for possible close encounter to the Solar System and found that on average our own Sun has 19.7$\pm$2.2 encounters within 1 parsec every million years. Therefore, studying disintegrating multiple systems can also provide us fundamental information on the risk of possible disruption to our own Solar System due to these close encounters.           

The recent discovery of candidates for free floating planets \citep{2020ApJ...903L..11M,2021MNRAS.505.5584M} begs the questions on how these objects come to exist. It is unlikely that they form on their own and the most likely explanation is that they are ejected from their own planetary systems \citep{1996Sci...274..954R}, but how these ejections happen is still an open question. One possibility is that they get ejected during the disintegration of a multiple system because it is likely that the planets orbiting the components get disrupted during the close interactions that lead to the disintegration of the systems \citep{2012MNRAS.422.1648V,2014MNRAS.437.1127V}.  
  
In this paper we aim to identify and study low-mass components ejected from post-encounter disintegrating multiple systems. These objects can be characterised more easily because they avoid the technological challenge of having to block the glare of the bright parent stars, a problem that has so far limited the characterisation of exoplanets. Well studied low-mass stars, brown dwarfs and giant planets are fundamental benchmark to constrain the formation theory and the atmospheric models for these very cold objects.   

We will explain in Section 2 how we selected our sample of candidate disintegrating systems. In Section 3 we outline the observing procedures and the methods that we used for the data reduction. The selection of additional ejected components is presented in Section 4. In Section 5 we discuss these systems and their ejected components, while in Section 6 we list our conclusion and lay out possible future work.

\section{Sample selection of candidate disintegrating systems}
 
To identify candidate disintegrating systems we used a method first described in \citet{2016csss.confE.137Y}, where we applied it to the \textit{Hipparcos} \citep{2007ASSL..350.....V} and \textit{Gliese-Jahrei\ss} catalogues \citep{1979A&AS...38..423G}. With the release of the Tycho-Gaia Astrometric Solution catalogue \citep[TGAS;][]{2015A&A...574A.115M}, we applied this method to extend our search for further studies. The method is described in details in this Section.

First, we searched for objects close in the sky and with common distance. We used TGAS because it provides very accurate PM and distance measurement for 2.5 million stars with a wide range of spectral types. 
  
We began by trimming the TGAS catalog and only kept stars with total PM $>$ 30 mas yr$^{-1}$ because we want systems/stars that can collide with each other with a significant relative PM, so that the components will disintegrate with significantly different PM that we will be able to measure. Given the typical distance of TGAS stars (30-100 pc) this PM cut corresponds to a tangential velocity of 5-15 km s$^{-1}$. This cut risks removing real disintegrating systems where one or more of the components have low tangential velocity (because their velocity vector is aligned with the velocity vector of the Sun). The release of more accurate parallaxes by Gaia DR3 and future Gaia data releases will allow us to lower this threshold or even remove it entirely during future searches. Then we kept only stars with PM/$\sigma_{PM}$ $>$ 10 in order to select only very accurate measurements, so that the relative PM of the components can be determined with high significance. We also cut at a distance of $\leq$ 300 pc (where we use distance = $1/\varpi$) because we want to identify ultra cool dwarfs (hereafter UCD) in these systems and given the intrinsic faintness of UCD we can only detect them out to $\sim$300 pc. The resulting sample had typical parallax uncertainties in TGAS of $<$10\%.   

The following step was to search systems out to a maximum separation. The widest known binary systems have separations of $\sim$200 kAU \citep[e.g.][]{2010A&A...514A..98C,2006A&A...460..635C} however the maximum separation observed for a main sequence star-brown dwarf binary is $\sim$5000 AU \citep{2006MNRAS.368.1281P,2001AJ....121.2185G} and white dwarf-brown dwarf binary have separations that can be up to 4 times larger \citep[i.e.$\sim$20 kAU, see e.g.][]{2011MNRAS.410..705D,2010MNRAS.404.1817Z,2010AJ....139..176F,2005AJ....129.2849B}. Overall, considering the fact that the systems we are searching for are disintegrating hence are expected to be extremely wide, we apply a conservative projected separation constraint of 50 kAU. This strikes a balance between the need to reduce the number of contaminants and our objective of selecting all possible candidates.

Afterwards, we removed from the sample all systems where the components are at a distance that is not consistent with each other within 3 times their TGAS uncertainties. Once we identify the initial group of candidates we have to also find the possible cause of the disintegration. As discussed in the introduction, this could be the dynamical interaction between the candidate and a nearby star or system. We therefore search for additional objects out to a 1 degree radius around our candidate systems. This separation cut is chosen as follows. First, we can only identify a disintegrating system as such if the disintegration has happened within the last few thousand years, otherwise the objects involved would have moved so far apart that a confident reconstruction of the events would be impossible. Therefore, if the disintegration has happened only a few thousand years ago, the component cannot have travelled further than 1 degree from each other even assuming the highest PM known for a star \citep[$\sim$10 arcsec yr$^{-1}$ for Barnard's Star;][]{2018A&A...616A...1G}.

We then needed to differentiate between gravitationally bound systems and unbound disintegrating systems. If the components are still bound, then they should have common PM (hereafter CPM). On the other hand disintegrating system should not have CPM. So we removed systems where the difference between the PM of all components are within the combined errors, and we keep groups where the PM of at least one object diverges from the others by at least three times the combined errors.

Finally we established whether the components of the systems were moving away from each other as expected for a disintegrating system or they were moving towards each other indicating a future close encounter. The separation between components as a function of time was computed as follow:

\begin{equation}
    \begin{aligned}
        {\rm Sep(t)} &= \sqrt{[(\alpha_1 + \mu_{\alpha_1}^* \times t) - (\alpha_2 + \mu_{\alpha_2}^* \times t)]^2 \times cos(\delta_1)^2} \\
        &\quad + \sqrt{[(\delta_1 + \mu_{\delta_1} \times t) - (\delta_2 + \mu_{\delta_2} \times t)]^2}
    \end{aligned}
    \label{eq:sep_t}
\end{equation}

Where $\alpha_1$, $\alpha_2$ and $\delta_1$, $\delta_2$ is the right ascension and the declination of the components in the system respectively, $\mu_{\alpha_1}^*$, $\mu_{\alpha_2}^*$ and $\mu_{\delta_1}$, $\mu_{\delta_2}$ is the PM in the right ascension direction and the PM in the declination direction of the components in the system respectively\footnote{In this paper, we use $\mu_\alpha^*$ to indicate that the PM in the R.A. direction is already multiplied by cos$\delta$.}.
    
Equation \ref{eq:sep_t} can be used to trace both the past and future motion of the system. However this equation assumes straight line motion for the components and ignores any gravitational attraction between them. This is a reasonable approximation for the majority of the systems because at wide separation the effect of gravity should be small. Nonetheless this approximation can lead to a small bias in the estimate of the time of closest approach and therefore we expect larger scatter in higher order systems where the complex gravitational interaction between the components add up. Using Equation \ref{eq:sep_t} we computed the time of closest encounter for all of our candidate systems. We propagated the uncertainties on the coordinate and PM using a Monte Carlo method. 

\begin{figure}
    \centering
    \includegraphics[width=\columnwidth, angle=180]{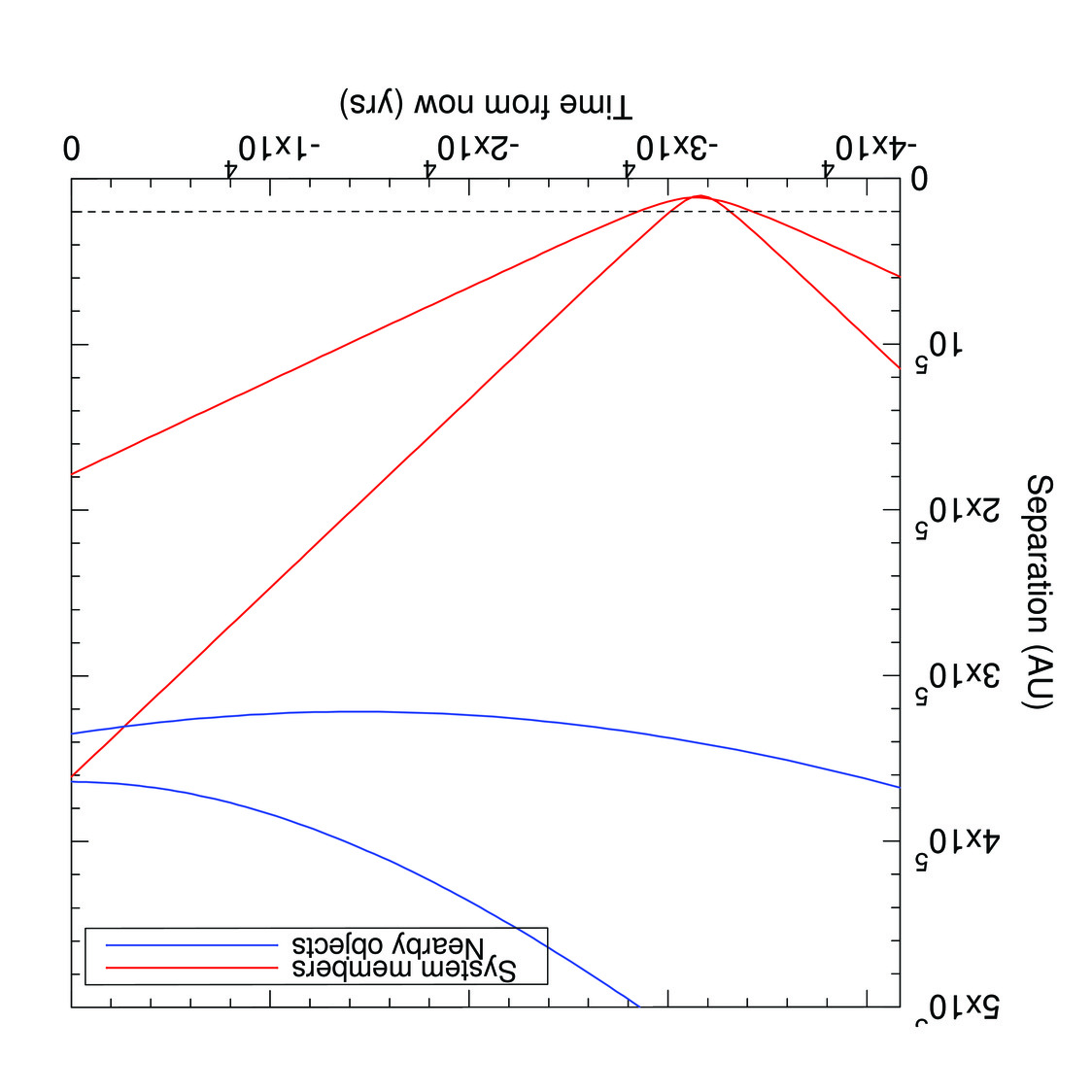}
    \caption{Back-tracked projected separation (AU) from TYC 4936-84-1, which is the central member of one of our candidate disintegrating system. The other two member of the system are plotted in red, while additional un-associated stars in the field are shown in blue. The dotted line indicates a projected separation of 20,000 AU which is a typical separation of a wide binary. All three component of the system where within this limit $\sim$32,000 years ago.}
    \label{fig:da_parabola}
\end{figure}

Finally we retain as our final disintegrating candidates only systems that have a time of closest encounter in the past, meaning that they are currently dispersing. For some systems some of the components had closest encounter in the past and some in the future. In those cases we visually inspected the plot of separation as a function of time and interpreted case by case whether the system was disintegrating or not. In Figure \ref{fig:da_parabola} we show an example of a disintegrating system and of additional un-associated objects in the field.  

Although the initial selection of candidates was done using TGAS, when Gaia DR3 \citep{2022arXiv220800211G} was released we updated all of the astrometric information for our systems. All of the results presented in this paper are based on this updated astrometry. 

Using the above method we identified 15 candidate disintegrating systems. In this paper, we present the results of our follow-up of 5 systems, which we observed to search for possible additional low-mass members. The 5 systems presented here are simply those that were observable from the Blanco Telescope during the night of 2017-04-22. They are listed in Table \ref{tab:da_astrometric} where for every object we give the Gaia DR3 coordinates, parallaxes, and PM. 

\begin{table*}
    \caption{Astrometric data for the candidate disintegrating systems presented here. The name is from SIMBAD, and a ``?'' next to the name indicates that the association of this object with the system in question is dubious (see Section 5 for further details). Gaia ID is the Gaia DR3 designation. R.A., Dec, PM ($\mu_\alpha^*$, $\mu_\delta$) and parallax ($\varpi$) are from Gaia DR3. Sep. is the current separation between the components. The time of closest encounter ($t_{\rm closest}$) is derived using equation \ref{eq:sep_t}.}
    \centering
    \begin{tabular}{l c c c c c c c c}
        \hline
        Name & Gaia ID & R.A. & Dec. & $\mu_\alpha^*$ & $\mu_\delta$ & $\varpi$ & Sep. & $t_{\rm closest}$ \\
             &      & (deg) & (deg) & (mas yr$^{-1}$) & (mas yr$^{-1}$) & (mas) & (arcsec) & (yr) \\
        \hline
        TYC 6813-1293-1 & 6046875561183621120 & 251.840515 & -24.706553 & -28.89$\pm$0.08 & 
        -34.89$\pm$0.06 & 8.42$\pm$0.07 & 0.0 & -29956.0 \\
        TYC 6813-286-1 & 6046740488740710016 & 251.183212 & -25.216904 & -107.2$\pm$0.3 &
        -98.5$\pm$0.2 & 9.8$\pm$0.2 & 2824.5 & -29956.0 \\
        TYC 6813-643-1A & 6046753549752392576 & 251.784424 & -25.158779 & -25.55$\pm$0.03 &
        -88.66$\pm$0.03 & 8.82$\pm$0.03 & 1638.3 & -29956.0 \\
        TYC 6813-643-1B & 6046753584102505856 & 251.784139 & -25.158305 & -28.3$\pm$0.1 &
        -87.59$\pm$0.07 & 8.98$\pm$0.09 & 1636.7 & -29956.0 \\
        \hline
        TYC 7240-1438-1 & 3462138227811420160 & 184.133972 & -34.44865 & -33.19$\pm$0.02 &
        -2.45$\pm$0.02 & 5.49$\pm$0.02 & 0.0 & -92590.0 \\
        TYC 7240-1159-1 & 3462913078568848256 & 183.732773 & -34.064308 & -47.13$\pm$0.03 &
        11.41$\pm$0.02 & 4.84$\pm$0.03 & 1827.4 & -92590.0 \\
        TYC 7240-850-1 & 3462739321372618112 & 182.755676 & -35.062286 & -73.6$\pm$0.02 &
        -23.96$\pm$0.01 & 5.64$\pm$0.02 & 4636.6 & -92590.0 \\
        \hline
        TYC 4936-84-1A & 3593854124477499520 & 174.488144 & -5.006651 & -49.37$\pm$0.05 & 
        -34.00$\pm$0.02 & 10.50$\pm$0.03 & 0.0 & -30233.0 \\
        TYC 4936-84-1B & 3593854124477587328 & 174.488539 & -5.006091 & -51.9$\pm$0.3 &
        -32.6$\pm$0.1 & 10.3$\pm$0.1 & 2.5 & -30233.0 \\
        TYC 4933-912-1A & 3792133076403723904 & 173.969406 & -4.017341 & -106.85$\pm$0.03 & 
        74.58$\pm$0.02 & 11.91$\pm$0.02 & 4018.7 & -30233.0 \\
        TYC 4933-912-1B & 3792133076402865792 & 173.968396 & -4.013940 & -106.00$\pm$0.02 & 
        74.05$\pm$0.01 & 11.89$\pm$0.01 & 4031.2 & -30233.0 \\
        TYC 4934-796-1 & 3599814851889741056 & 175.038681 & -4.947755 & 15.44$\pm$0.02 & 
        -23.85$\pm$0.02 & 10.13$\pm$0.02 & 1985.8 & -30233.0 \\       
        \hline
        TYC 9281-3037-1 & 5806792421244550656 & 241.931213 & -71.349174 & -367.21$\pm$0.02 & 
        288.76$\pm$0.03 & 6.11$\pm$0.03 & 0.0 & -9677.0 \\
        TYC 9281-2422-1 & 5806506204622362112 & 244.929993 & -72.214973 & -20.04$\pm$0.05 & 
        -43.30$\pm$0.06 & 5.18$\pm$0.06 & 4593.2 & -9677.0 \\
        TYC 9281-1175-1A & 5806506685658712832 & 244.605225 & -72.236076 & -46.59$\pm$0.01 & 
        -31.74$\pm$0.01 & 5.46$\pm$0.01 & 4385.7 & -9677.0 \\
        TYC 9281-1175-1B? & 5806505208191451008 & 244.606052 & -72.236789 & \ldots & 
        \ldots & \ldots & 4388.2 & -9677.0 \\        
        \hline 
        TYC 7731-1951-1 & 5391597005019395200 & 162.216339 & -41.780659 & -95.21$\pm$0.05 &
        10.88$\pm$0.08 & 6.32$\pm$0.07 & 0.0 & -8663.0 \\
        TYC 7731-2128-1A & 5391592538253547648 & 162.319901 & -41.896751 & -56.518$\pm$0.009 &
        -31.27$\pm$0.01 & 4.35$\pm$0.01 & 501.8 & -8663.0 \\
        TYC 7731-2128-1B & 5391592538254935296 & 162.319848 & -41.896005 & -57.34$\pm$0.09 &
        -31.5$\pm$0.1 & 4.4$\pm$0.1 & 499.5 & -8663.0 \\
        TYC 7731-1995-1AB? & 5391594668557342592 & 162.366165 & -41.806747 & -36.77$\pm$0.01 
        & 3.08$\pm$0.02 & 4.04$\pm$0.02 & 413.0 & -8663.0 \\
        TYC 7731-1995-1C? & 5391594668558722048 & 162.367068 & -41.806130 & \ldots 
        & \ldots & \ldots & 414.8 & -8663.0 \\
        \hline            
    \end{tabular}
    \label{tab:da_astrometric}
\end{table*}

Next we calculated the UVW component of the Galactic velocity and the XYZ position inside the Galaxy for all of the members of our disintegrating systems. We did this using the Gaia DR3 position and PM and the radial velocity (hereafter RV) from Gaia or from the literature. We only found RV measurements for 16 out of 21 of the stars in the 5 systems studied here. The calculations were done using the IDL program GAL\_UVW from the astronomy users library and our own IDL code.

To calculate the mass for the stellar component of our systems we used a mass-luminosity relation. First we calculate the absolute $V$ magnitude for each star using the $V$ magnitude from TYCHO and the parallax from Gaia DR3. Then for F, G and K dwarfs we interpolated the updated version of Table 5 from \citet{2013ApJS..208....9P} and Table 3 of \citet{2012ApJ...746..154P}, which is available at \url{http://www.pas.rochester.edu/~emamajek/EEM_dwarf_UBVIJHK_colors_Teff.txt}. For the M dwarfs we used the mass-luminosity relation of equation 11 from \citet{2016AJ....152..141B} with the coefficients from their Table 13. 

Finally we searched through the literature as well as through large area survey such as the GALactic Archaeology with HERMES survey \citep[GALAH;][]{2021MNRAS.506..150B} and The RAdial Velocity Experiment \citep[RAVE;][]{2017AJ....153...75K} to find metallicity values for our stars. We found metallicity values for 16 out of 21 stars in our sample. When comparing [Fe/H] measurements from different surveys it is important to take into account possible systematic offsets between surveys. The metallicity for 13 out of 16 of our objects come from three surveys: GALAH (5 objects), RAVE (4 objects), and Gaia DR3 (4 objects). The exceptions are TYC 6813-286-1 for which we get the metallicity from \citet{2006AeA...454..895A}, and TYC 9281-2422-1 and TYC 7731-2128-1A for which we get the metallicity from \citet{2006ApJ...638.1004A}. GALAH validated their metallicity estimates using the Gaia FGK Benchmark Stars \citep[Version 2.1;][]{2018RNAAS...2..152J} and found a systematic [Fe/H] shift of +0.1 which they applied to their published metallicity values \citep[see section 4.1.3 in][]{2021MNRAS.506..150B}. RAVE performed an extensive comparison of their metallicities with several other surveys, including GALAH. \citet{2017AJ....153...75K} found negligible [Fe/H] systematic offsets between RAVE and GALAH of $ -0.07\pm0.45$ for stars with signal-to-noise ratio $ <50$ and $ +0.04\pm0.13$ for stars with signal-to-noise ratio $ >50$ \citep[see Section 7.5 and Table 5 in][]{2017AJ....153...75K}. Gaia DR3 also checked the accuracy of their metallicity estimates against a number of surveys, including GALAH and RAVE. \citet{2022arXiv220605992F} found that the GSP-Phot metallicities from Gaia are typically too low by 0.2 and have a median absolute deviation with respect to the other surveys of 0.2 \citep[see Section 3.2.1 in][]{2022arXiv220605992F}. In this paper we have applied the 0.2 systematic correction to the values we took from Gaia, and we use $ \pm0.2$ as their uncertainty. Finally, we found no systematic comparison between the [Fe/H] published by \citet{2006AeA...454..895A} and \citet{2006ApJ...638.1004A} and other surveys, so we cannot comment on possible systematic offsets in those values. Overall, we conclude that there are no remaining systematic offsets in the [Fe/H] values we use here, with the possible exception of the [Fe/H] values for TYC 6813-286-1, TYC 9281-2422-1, and TYC 7731-2128-1A.

Spectral type, mass, metallicity, RV, and UVWXYZ are given in Table \ref{tab:da_properties}. The reference for each [Fe/H] and RV value is listed in the table.

\begin{landscape} 
\begin{table}
\scriptsize
    \caption{Properties for the stellar components of our disintegrating multiple systems. The spectral type is taken from the literature, the mass was estimated using the method described in Section 2. References for RV and metallicity are given in the table. UVWXYZ are calculated using the Gaia DR3 parallaxes, PM, and the RV given in the table.}
    \begin{tabular}{l c c c c c c c c c c c c}
        \hline
        Name & SpT & Mass & [Fe/H] & [Fe/H] ref. & RV & RV ref. & U & V & W & X & Y & Z \\
             &  & ($M_\odot$) & (dex) &  & (km s$^{-1}$) &  & (km s$^{-1}$) & (km s$^{-1}$) & (km s$^{-1}$) & (pc) & (pc) & (pc) \\
        \hline
        TYC 6813-1293-1 & F5V & 1.14$\pm$0.05 & -0.05$\pm$0.2 & \citealt{2022arXiv220800211G}
        & -46.6$\pm$0.3 & \citealt{2018AeA...616A...1G} & 47.1$\pm$0.3 & -22.9$\pm$0.2 & -9.95$\pm$0.08 & -33.7$\pm$0.3 & -104.1$\pm$0.9 & -49.7$\pm$0.4 \\
        TYC 6813-286-1 & F7V+A(pSr) & 1.14$\pm$0.05 & -0.11$\pm$0.04 & \citealt{2006AeA...454..895A}
        & -69.0$\pm$5.0 & \citealt{2018AeA...616A...1G} & 75.0$\pm$4.0 & -66.0$\pm$1.0 & -5.0$\pm$1.0 & -29.2$\pm$0.6 & -90.0$\pm$2.0 & -43.0$\pm$0.9 \\
        TYC 6813-643-1A & K0IV & 0.85$\pm$0.05 & -0.1$\pm$0.05 & \citealt{2021MNRAS.506..150B}
        & -61.2$\pm$0.3 & \citealt{2021MNRAS.506..150B} & 58.2$\pm$0.3 & -43.0$\pm$0.2 & -32.49$\pm$0.09 & -32.4$\pm$0.1 & -99.8$\pm$0.3 & -47.7$\pm$0.2 \\
        TYC 6813-643-1B & M3.5V & \ldots & \ldots & \ldots
        & \ldots & \ldots & \ldots & \ldots & \ldots & \ldots & \ldots & \ldots \\
        \hline
        TYC 7240-1438-1 & F3V & 1.25$\pm$0.05 & 0.0$\pm$0.2 & \citealt{2022arXiv220800211G}
        & 12.4$\pm$0.3 & \citealt{2022arXiv220800211G} & 19.6$\pm$0.1 & -24.4$\pm$0.2 & 0.1$\pm$0.1 & -148.8$\pm$0.5 & -12.98$\pm$0.05 & -102.2$\pm$0.4 \\
        TYC 7240-1159-1 & F8V & 1.08$\pm$0.05 & 0.05$\pm$0.06 & \citealt{2017AJ....153...75K}
        & -9.4$\pm$0.5 & \citealt{2022arXiv220800211G} & 47.3$\pm$0.3 & -10.4$\pm$0.4 & -1.2$\pm$0.2 & -171.0$\pm$1.0 & -14.77$\pm$0.09 & -117.1$\pm$0.7 \\
        TYC 7240-850-1 & G5V & 1.08$\pm$0.05 & 0.07$\pm$0.07 & \citealt{2017AJ....153...75K}
        & 20.7$\pm$0.2 & \citealt{2022arXiv220800211G} & 39.6$\pm$0.2 & -52.7$\pm$0.2 & -17.9$\pm$0.1 & -144.2$\pm$0.5 & -11.72$\pm$0.04 & -99.2$\pm$0.4 \\
        \hline
        TYC 4936-84-1A & K1III & 0.82$\pm$0.05 & -0.12$\pm$0.09 & \citealt{2017AJ....153...75K}
        & -1.8$\pm$0.8 & \citealt{2018AeA...616A...1G} & 12.11$\pm$0.04 & -18.6$\pm$0.5 & -15.6$\pm$0.7 & -94.6$\pm$0.3 & 8.44$\pm$0.02 & -7.84$\pm$0.02 \\
        TYC 4936-84-1B & M4.5V & \ldots & \ldots & \ldots & \ldots & \ldots & \ldots & \ldots & \ldots & \ldots & \ldots & \ldots \\
        TYC 4933-912-1A & K0V & 0.87$\pm$0.05 & 0.19$\pm$0.05 & \citealt{2021MNRAS.506..150B}
        & 15.90$\pm$0.07 & \citealt{2021MNRAS.506..150B} & 51.03$\pm$0.09 & -5.55$\pm$0.04 & 16.64$\pm$0.06 & -82.9$\pm$0.1 & 7.44$\pm$0.01 & -6.76$\pm$0.01 \\
        TYC 4933-912-1B & K8V & 0.70$\pm$0.05 & 0.02$\pm$0.07 & \citealt{2017AJ....153...75K}
        & 17.3$\pm$0.8 & \citealt{2022arXiv220800211G} & 51.33$\pm$0.04 & -5.8$\pm$0.5 & 17.0$\pm$0.6 & -82.96$\pm$0.08 & 7.465$\pm$0.007 & -6.754$\pm$0.006 \\
        TYC 4934-796-1 & K0V & 0.87$\pm$0.05 & 0.29$\pm$0.05 & \citealt{2021MNRAS.506..150B}
        & -19.89$\pm$0.06 & \citealt{2021MNRAS.506..150B} & -11.45$\pm$0.02 & 6.80$\pm$0.04 & -19.93$\pm$0.05 & -98.6$\pm$0.2 & 8.78$\pm$0.02 & -8.22$\pm$0.02 \\ 
        \hline
        TYC 9281-3037-1 & G0V & 1.08$\pm$0.05 & -2.3$\pm$0.1 & \citealt{2021MNRAS.506..150B} 
        & 238.1$\pm$0.9 & \citealt{2021MNRAS.506..150B} & -206.2$\pm$0.7 & -250.6$\pm$0.8 & 281.0$\pm$2.0 & -22.5$\pm$0.1 & -44.1$\pm$0.2 & -153.5$\pm$0.8 \\
        TYC 9281-2422-1 & G5V & 1.07$\pm$0.05 & 0.2$\pm$0.4 & \citealt{2006ApJ...638.1004A}
        & 16.0$\pm$6.0 & \citealt{2022arXiv220800211G} & 19.4$\pm$4.3 & -38.4$\pm$3.9 & -17.6$\pm$1.6 & -25.1$\pm$0.3 & -53.7$\pm$0.6 & -183.0$\pm$2.0 \\
        TYC 9281-1175-1A & F8V & 1.14$\pm$0.05 & 0.06$\pm$0.08 & \citealt{2021MNRAS.506..150B}
        & -5.2$\pm$0.1 & \citealt{2021MNRAS.506..150B} & 34.02$\pm$0.09 & -33.70$\pm$0.09 & 11.44$\pm$0.03 & -23.97$\pm$0.04 & -50.78$\pm$0.09 & -174.4$\pm$0.3 \\
        TYC 9281-1175-1B? & M5V & \ldots & \ldots & \ldots
        & \ldots & \ldots & \ldots & \ldots & \ldots & \ldots & \ldots & \ldots \\ 
        \hline
        TYC 7731-1951-1 & G5V & 1.08$\pm$0.05 & -0.3$\pm$0.2 & \citealt{2022arXiv220800211G}
        & 32.6$\pm$0.3 & \citealt{2022arXiv220800211G} & 59.4$\pm$0.7 & -49.2$\pm$0.3 & -16.7$\pm$0.3 & -112.0$\pm$1.0 & 35.5$\pm$0.4 & -105.0$\pm$1.0 \\
        TYC 7731-2128-1A & G0V & 1.21$\pm$0.05 & 0.3$\pm$0.3 & \citealt{2006ApJ...638.1004A}
        & 0.6$\pm$0.2 & \citealt{2022arXiv220800211G} & 35.1$\pm$0.1 & -22.6$\pm$0.2 & -56.7$\pm$0.1 & -163.3$\pm$0.4 & 51.6$\pm$0.1 & -153.3$\pm$0.4 \\
        TYC 7731-2128-1B & M3V & \ldots & \ldots & \ldots
        & \ldots & \ldots & \ldots & \ldots & \ldots & \ldots & \ldots & \ldots \\
        TYC 7731-1995-1AB? & K0III & 2.30$\pm$0.05 & 0.2$\pm$0.2 & \citealt{2022arXiv220800211G}
        & 5.3$\pm$0.1 & \citealt{2018AeA...616A...1G} & 37.6$\pm$0.2 & -16.3$\pm$0.1 & -14.93$\pm$0.09 & -175.9$\pm$0.9 & 55.6$\pm$0.3 & -165.0$\pm$0.8 \\
        TYC 7731-1995-1C? & M3V & \ldots & \ldots & \ldots
        & \ldots & \ldots & \ldots & \ldots & \ldots & \ldots & \ldots & \ldots \\
        \hline
    \end{tabular}
    \label{tab:da_properties}
\end{table}
\end{landscape} 

\section{Observations and data reduction}

To search for additional UCD components, too cold and faint to be seen by Gaia, ejected from our disintegrating multiple systems we obtained deep imaging of a large area around each system with the Dark Energy Camera \citep[DECam;][]{2015AJ....150..150F} on the Blanco 4m Telescope. We observed each system with the $r$, $i$, $z$ and $Y$ filters obtaining 7 exposures on a dithered pattern with offset of 60 arcsecs. The total exposure time for each filter is given in Table \ref{tab:da_FilterExposureTime}. The exposure times were chosen to reach a S/N=10 depth of $r$$\sim$24 mag, $i$$\sim$24 mag, $z$$\sim$22 mag and $Y$$\sim$21 mag. For calibration we obtained dome flats, biases and darks. We observed 3 photometric standard fields at different airmass for photometric calibration: SDSSJ1048-0000, SDSSJ0933-0005 and SDSSJ0843-0000. 

The images were processed using the DECam Community Pipeline \citep{2014ASPC..485..379V}, while the photometry was obtained from the images with our own pipeline based on the PSF-fitting algorithm \textsc{daophotII/ALLSTAR} \citep{1987PASP...99..191S}. The final catalog includes only stellar-shaped objects with |\emph{sharpness}| $\leq$ 0.5 to avoid, as much as possible, the presence of non-stellar sources and background galaxies in our analysis. 

We converted the instrumental magnitude to calibrated magnitude using the following equation: 

\begin{equation}
    \begin{aligned}
        r({\rm AB}) =& 0.998170(\pm0.000067) \times M_{\rm apcor} - 0.04394(\pm0.00029) \times \\
              &{\rm Airmass} + 0.37797
    \end{aligned}
\end{equation}

\begin{equation}
    \begin{aligned}
        i({\rm AB}) =& 0.996426(\pm0.000071) \times M_{\rm apcor} - 0.03293(\pm0.00030) \times \\
              &{\rm Airmass} + 0.31727
    \end{aligned}
\end{equation}

\begin{equation}
    \begin{aligned}
        z({\rm AB}) =& 0.99785(\pm0.00012) \times M_{\rm apcor} - 0.04291(\pm0.00041) \times \\ 
              &{\rm Airmass} - 0.07681
    \end{aligned}
\end{equation}

\begin{equation}
    \begin{aligned}
        Y({\rm AB}) =& 1.016258(\pm0.00035) \times M_{\rm apcor} - 0.03754(\pm0.00010) \times \\
              &{\rm Airmass} - 1.479324
    \end{aligned}
\end{equation}

Where $M_{\rm apcor}$ is the aperture-corrected instrumental magnitude. 

We also calculated PM for the stars in our DECam images. To do that we crossed-matched our DECam observations with the Visible and Infrared Survey Telescope for Astronomy (VISTA) Hemisphere Survey \citep[VHS;][]{2021yCat.2367....0M}, the United Kingdom Infrared Telescope (UKIRT) Infrared Deep Sky Survey \citep[UKIDSS;][]{2007MNRAS.379.1599L} and the Two Micron All Sky Survey \citep[2MASS;][]{2006AJ....131.1163S}. We calibrated the measured position for our DECam objects using Gaia DR2 as a reference. First we cross-matched the DECam observations with Gaia DR2 with a radius of 5 arcsec and we kept only Gaia matches that have a measured PM. Then we used the Gaia PM to move each Gaia reference star to the epoch of our DECam observations. We then calculated adjustment to the World Co-ordinate System (WCS) of the DECam images using a least square fit with 3$\sigma$ outlier rejection. Finally we measured the PM with a linear fit to the positions in the available data, i.e. combining our DECam epoch with the VHS, UKIDSS, and 2MASS data.  

\begin{table*}
    \caption{Observation details for the systems presented in this paper. For each system we listed the name of the components, the filter used and the exposure time for each filter.}
    \centering
    \begin{tabular}{l c c}
        \hline
        Name & Filter & Exposure time \\
             &        & (s) \\
        \hline
        TYC 6813-1293-1, TYC 6813-286-1, TYC 6813-643-1AB & $r$ & 180\,$\times$\,7  \\ 
                                                          & $i$ & 180\,$\times$\,7  \\
                                                          & $z$ & 50\,$\times$\,7  \\
                                                          & $Y$ & 40\,$\times$\,7  \\
        \hline
        TYC 7240-1438-1, TYC 7240-1159-1, TYC 7240-850-1 & $r$ & 180\,$\times$\,7  \\ 
                                                         & $i$ & 180\,$\times$\,7  \\
                                                         & $z$ & 50\,$\times$\,7  \\
                                                         & $Y$ & 40\,$\times$\,7  \\
        \hline
        TYC 4936-84-1AB, TYC 4933-912-1AB, TYC 4934-796-1 & $r$ & 180\,$\times$\,7  \\ 
                                                          & $i$ & 180\,$\times$\,7  \\
                                                          & $z$ & 50\,$\times$\,7  \\
                                                          & $Y$ & 40\,$\times$\,7  \\
        \hline
        TYC 9281-3037-1, TYC 9281-2422-1, TYC 9281-1175-1AB? & $r$ & 180\,$\times$\,7  \\
                                                            & $i$ & 180\,$\times$\,7  \\
                                                            & $z$ & 50\,$\times$\,7  \\
                                                            & $Y$ & 40\,$\times$\,7  \\
        \hline
        TYC 7731-1951-1, TYC 7731-2128-1AB, TYC 7731-1995-1ABC? & $r$ & 180\,$\times$\,7  \\ 
                                                               & $i$ & 180\,$\times$\,7  \\
                                                               & $z$ & 50$\,\times$\,7  \\
                                                               & $Y$ & 40$\,\times$\,7  \\
        \hline
    \end{tabular}
    \label{tab:da_FilterExposureTime}
\end{table*}

\section{Selection of additional low-mass ejected component}

We search for additional low-mass component of our disintegrating multiple systems using the photometry and PM measured with our DECam data (see Section 3 for details). First we used the PMs to calculate the distance that each object in the DECam images had from the centre of the disintegrating system at the time of closest encounter (i.e. column 8 in Table \ref{tab:da_astrometric}). We selected only objects that were within 20,000 AU of the centre of the system, since this is the separation of the widest binaries containing UCDs known to date (see Section 2). We calculated the uncertainty on the separation as a function of time by propagating the uncertainty on the measured position and PM using a Monte Carlo method as follows. For each object, we generate 10,000 ``copies'' with coordinates and PM taken from a Gaussian distribution centered on the measured values, and with $\sigma$ equal to the measurement uncertainties. We then computed the separation as a function of time for all of those 10,000 ``copies''. We assumed as uncertainty the standard deviation of the distribution of 10,000 values. We then looked at the direction of the motion for all of the objects and compared with the direction of the motion of the stellar components of the system. We retained only those objects where the direction of motion is consistent with having been ejected after the close encounter. 

Next we used the DECam photometry and any additional NIR magnitude from the literature to construct a series of colour-magnitude diagrams (hereafter CMD) and used those to select UCD candidates. We used the CMDs from \citet{2018ApJS..234....1B}, \citet{2021AJ....161...42B}, and \citet{2011ApJS..197...19K} as reference to estimate the spectral type of our candidates based on their colours. We also derived spectral type estimates using the colour to spectral type and absolute magnitude to spectral type relations from \citet{2021ApJS..253....7K}, estimating the absolute magnitude of the UCD candidates by assuming that they were at the same distance as the stars in the disintegrating systems. Finally, we also estimated spectral types for the UCD candidates by comparing their colours with the reference colours from Table 1 of \citet{2015A&A...574A..78S}. Objects with consistent spectral types from all methods are our most promising candidates. Given the uncertainties in the measured magnitudes and the large intrinsic scatter among the population of UCDs, we keep objects with spectral types consistent within $\sim$4 subtypes. 

\begin{figure}
    \centering
    \includegraphics[width=\columnwidth]{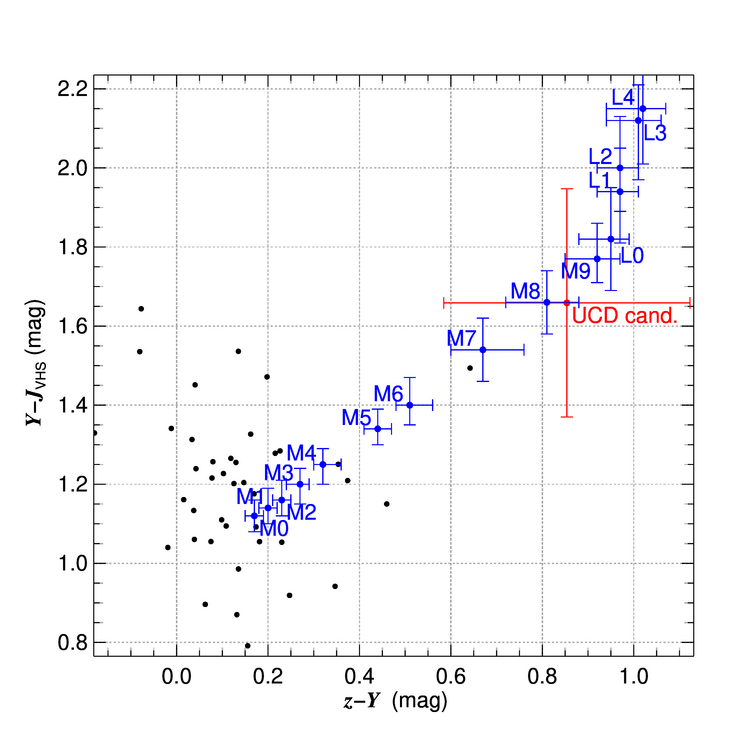}
    \includegraphics[width=\columnwidth]{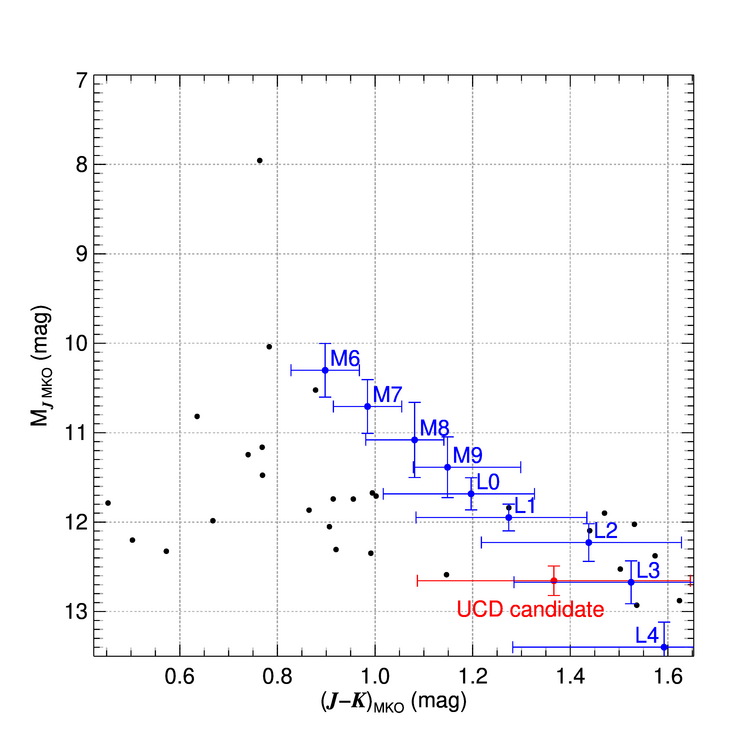}
    \caption{Colour-colour diagram (top panel) and colour-magnitude diagram (bottom panel) for the candidate additional components of the system composed of TYC 7731-1951-1, TYC 7731-2128-1AB, and TYC 7731-1995-1ABC?. The more promising candidate is highlighted in red, whereas initial candidates that were subsequently rejected because of their inconsistent colours and magnitudes are plotted in black. To improve the clarity of the Figure, we omit the error bars on rejected candidates. The uncertainty on the absolute magnitude of the UCD candidate is based on the $J$ magnitude uncertainty and the parallax error of the primary, i.e. TYC 7731-2128-1AB. In blue we mark the median colours and the 1-$\sigma$ scatter listed in \citet{2018ApJS..234....1B}. The $z$ and $Y$ magnitude are from our DECam observation, the $J$ magnitude is from the VHS.}
    \label{fig:system8_object85_CMD}
\end{figure}

With the analysis above we identified one promising UCD candidate associated with the disintegrating system consisting of TYC 7731-1951-1, TYC 7731-2128-1 AB, and TYC 7731-1995-1ABC?. Figure~\ref{fig:system8_object85_CMD} shows two examples of colour-colour and colour-magnitude diagrams used for the selection. The top panel shows the $z$-$Y$ vs $Y$-$J$ colour-colour diagram, while the bottom panel shows the $M_{J}$ vs $J$-$K$ colour-magnitude diagram. We highlight in red the position of our most promising candidate UCD. In both plots, its colours and magnitude are consistent with the location of late-M and early-L dwarfs, and the spectral types estimates we get from the \citet{2021ApJS..253....7K} polynomials and the \citet{2015A&A...574A..78S} colours are all consistent with this object having a spectral type in the range M8--L2. We discuss further this object and the system that contains it in Section 5. 

All other UCD candidates in the disintegrating systems presented here are discarded either because their colours and magnitudes lead to inconsistent spectral types, or because visual inspection of their PM and separation as a function of time are inconsistent with them being associated with the main sequence stars in the system.

\section{Discussion on individual systems}

\subsection{TYC 6813-1293-1, TYC 6813-286-1 and TYC 6813-643-1AB}

A very interesting system in our sample is the one that contains TYC 6813-1293-1, TYC 6813-286-1, and TYC 6813-643-1AB. 

TYC 6813-1293-1 is in the catalog of accelerating stars \citep{2021ApJS..254...42B}, which means that it is probably an unresolved binary. Its spectral type is F5 \citep{1988mcts.book.....H}, and it has a metallicity approximately 2 times higher than the other two stars in the system ([Fe/H]=-0.05$\pm$0.2; \citealt{2022arXiv220800211G}). Its RV is also smaller than the other two objects (RV=-46.6$\pm$0.3 km s$^{-1}$) \citep{2018A&A...616A...1G}. It has a probability of 99\% to be in the thin disk. The probability was calculated by first computing the UVW velocity for our targets using the Gaia PM, parallax \citep{2021A&A...649A...1G} and RV \citep{2018A&A...616A...1G}, and then assuming that the three components of the Galaxy (i.e. thin disk, thick disk, and halo) have gaussian UVW velocity distributions. The $\sigma$ of the three distributions was taken from \citep{2003A&A...410..527B}. The probability to belong to each of the three components is then calculated as follows:

\begin{equation}
    k_i = \frac{1}{(2\pi)^{3/2} \sigma_{Ui} \sigma_{Vi} \sigma_{Wi}}
\end{equation}
\begin{equation}
    f(U, V, W)_i = k_i\ exp\left(-\frac{U^2}{2\sigma^2_{Ui}} - \frac{(V - V_{asym_i})^2}{2\sigma^2_{Vi}} - \frac{W^2}{2\sigma_{Wi}^2}\right)
\end{equation}
\begin{equation}
    Prob_i = \frac{X_i\ f_i}{X_{\rm thin}\ f_{\rm thin} + X_{\rm thick}\ f_{\rm thick} + X_{\rm halo}\ f_{\rm halo}}
\end{equation}

\noindent where the subscript $i$ indicates either the thin disk, thick disk, or halo, $\sigma_{Ui}$, $\sigma_{Vi}$, and $\sigma_{Wi}$ are the velocity dispersion for each component of the Galaxy from \citet{2003A&A...410..527B}, and $V_{asym_i}$ is the asymmetric velocity drift for the same component, also taken from \citet{2003A&A...410..527B}, and $X_i$ is the fraction of stars belonging to each component as estimated from the Solar neighborhood population \citep[also taken from][]{2003A&A...410..527B}. The assumption that the UVW distributions are Gaussian is an approximation. For example we could model the thin disk distribution better using a combination of several Gaussian components with dispersion that depends on the age (as it is done for example in the Gaia Universe Model Snapshot; \citealt{2012A&A...543A.100R}). However, we believe that our approximation causes only small underestimations of the probability of an object to belong to the thin disk.

We also checked the probability of the target being in a moving group using BANYAN \citep{2018ApJ...856...23G}. It has a probability of 99.9\% to be a field object. 

TYC 6813-286-1 is also known to be a spectroscopic binary \citep{1988mcts.book.....H}. Its spectral type is F7+A(pSr) \citep{1988mcts.book.....H}. Its metallicity is 2 times lower than TYC 6813-1293-1 but almost the same as the other object in this system ([Fe/H]=-0.11$\pm$0.04) \citep{2006AeA...454..895A}. Its RV is also very similar to the other object in the system (RV=-69$\pm$5 km s$^{-1}$; \citealt{2018A&A...616A...1G}). The RV error is pretty big, which we speculate is due to TYC 6813-286-1 being an unresolved binary. It has a probability of 88\% to be in the thin disk, and a probability of 99.9\% to be in the field according to BANYAN \citep{2018ApJ...856...23G}. 

TYC 6813-643-1 is resolved by Gaia in two components which form a common-proper-motion-pair. The spectral type for the primary (hereafter A) is K0IV \citep{2010PASP..122.1437P}, and we estimate the spectral type of the companion (hereafter B) to be M3.5V using the Gaia magnitude and parallax and using Table 5 from \citet{2013ApJS..208....9P}\footnote{\url{http://www.pas.rochester.edu/~emamajek/EEM_dwarf_UBVIJHK_colors_Teff.txt}}. The RV and metallicity of A are very similar to those of TYC 6813-286-1, but with a smaller uncertainty -- RV=-61.2$\pm$0.3 km s$^{-1}$ \citep{2021MNRAS.506..150B}, [Fe/H]=-0.10$\pm$0.05 \citep{2021MNRAS.506..150B}. There is no RV and no metallicity measurement for B. A has a probability of 95\% to be in the thin disk, and a probability of 99.9\% of being a field object according to BANYAN \citep{2018ApJ...856...23G}. Because we don't have RV for B we cannot calculate those probabilities for it.

\begin{figure*}
    \centering
    \includegraphics[width=\columnwidth]{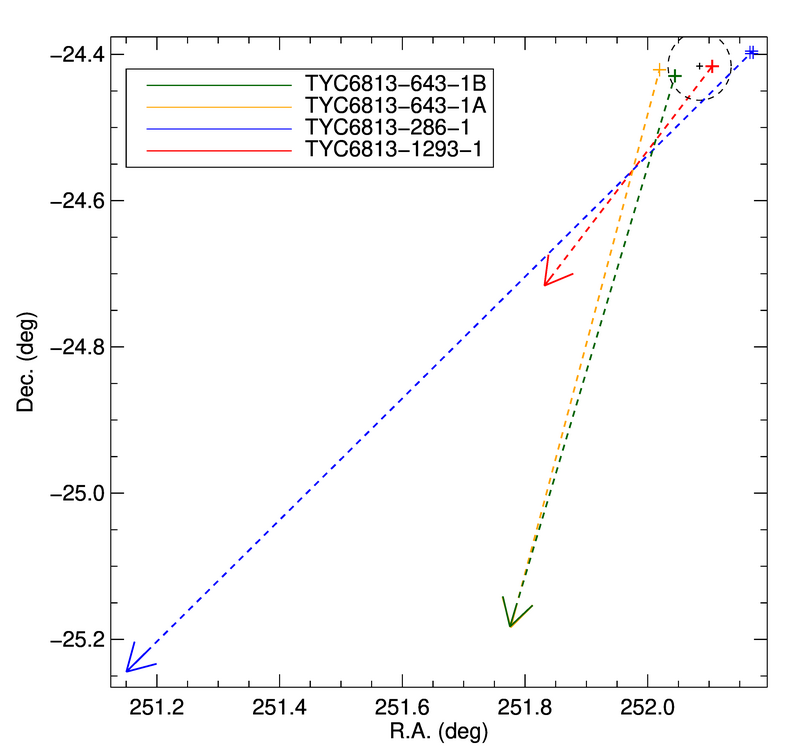}
    \includegraphics[width=\columnwidth]{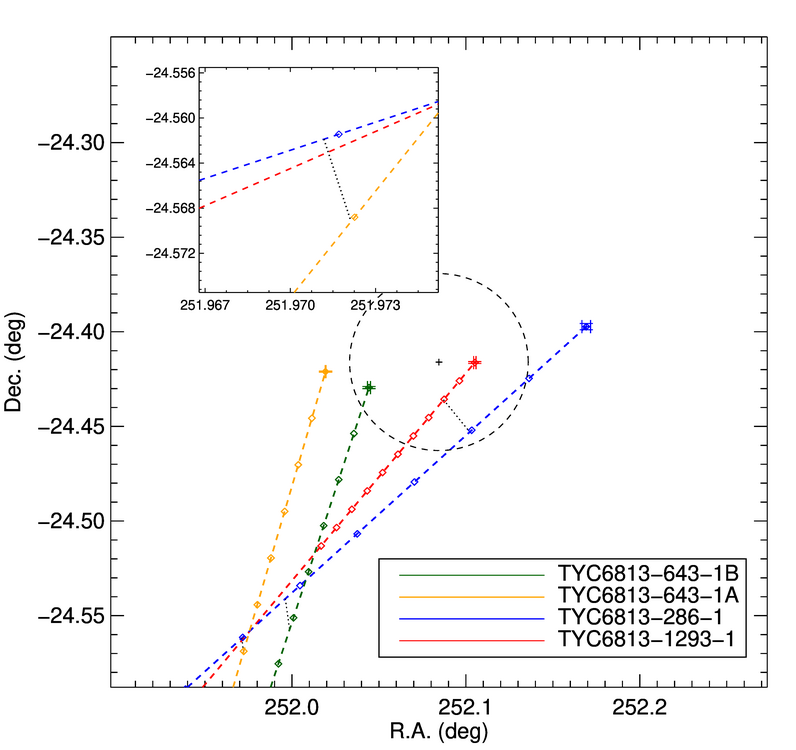}
    \includegraphics[width=\columnwidth]{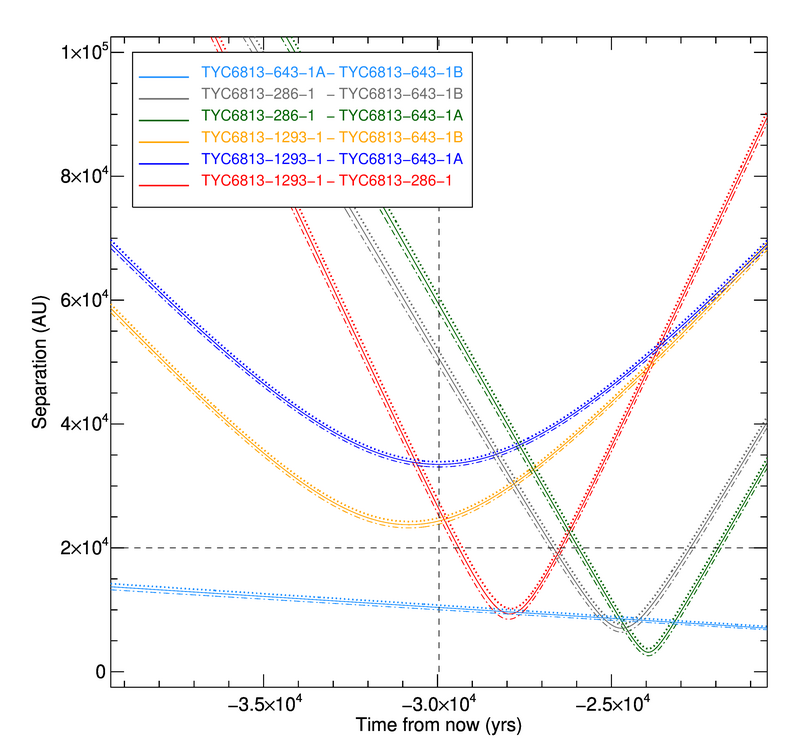}
    \begin{minipage}[b]{\columnwidth}
    \caption{\textit{Top left panel:} the position of TYC 6813-1293-1, TYC 6813-286-1 and TYC 6813-643-1 AB now and at the time of closest encounter of the system. For each object, the arrow indicates the current direction of motion, the dashed line connects the current position with the position at the time of closest encounter, marked by a cross. The small black cross is the centre of the system at the time of closest encounter, and the black dashed circle has a radius of 20,000 AU. TYC 6813-1293-1 passes between the other three objects, likely causing their disintegration. \textit{Top right panel:} same as the top left panel, but zoomed in on the position of closest encounter. Symbols and colours are the same as the top left panel, and we added diamonds to show the position of each object in intervals of 1,000 years after the time of closest encounter, up to a maximum of 10,000 years. For pairs of objects that did not reach their mutual minimum separation at the time of closest encounter for the system, a dark grey dotted line shows the position and the time when their interaction took place. TYC6813-1293-1 (red dashed line) is likely causing the disintegration of TYC6813-286-1 (blue dashed line) and TYC6813-643-1AB (orange and dark green dashed lines) by first interacting with TYC6813-643-1AB, and later interacting with TYC6813-286-1. The interactions between TYC6813-643-1B and TYC6813-286-1 and between TYC6813-643-1A and TYC6813-286-1 happened later on. \textit{Bottom left panel:} The separation between the components of the system as a function of time. The vertical dashed line marks the time of closest encounter while the horizontal dashed lines indicates a separation of 20,000 AU. The dotted and dash-dotted line above and below each parabola indicate the one-sigma uncertainty range. TYC 6813-1293-1 comes to a closer separation from TYC 6813-286-1 and TYC 6813-643-1AB than the separation between TYC 6813-286-1 and TYC 6813-643-1 AB (i.e. the red parabola is at smaller separation than the dark grey and dark green parabolas at the time of closest encounter), causing the breakup of the system. TYC 6813-643-1 AB, however, survives the encounter and continues onward as a binary (i.e. the dodger blue parabola, see Section 5.1 for discussion of its slope). The dark grey and dark green parabola do not reach the minimum at the same time, contrary to what one might expect from a binary. However, we think this is due to orbital motion for the binary. \label{fig:da_System2_arrow_parabola}}
    \end{minipage}
\end{figure*}

So, overall we assume that TYC 6813-286-1 and TYC 6813-643-1, each of which is a binary, used to form a quadruple system, as confirmed by their nearly identical RV and metallicity. Then they had a close encounter with TYC 6813-1293-1, which did not belong to the original system as demonstrated by its very different RV and metallicity. Now the quadruple is disintegrating. Figure~\ref{fig:da_System2_arrow_parabola} shows this more clearly. As one can see in the top left and top right panels, TYC 6813-1293-1 (the red arrow and dashed line) passes between the other three objects involved during the close encounter causing the disintegration. In particular, the top right panel of Figure~\ref{fig:da_System2_arrow_parabola} shows that the interaction between TYC6813-1293-1 and TYC6813-643A and B happens first, followed by the interaction between TYC6813-1293-1 and TYC6813-286-1 (approximately 2,000 years after the system closest encounter). TYC6813-286-1 and TYC6813-643-1A and B interact with each other approximately 5,000-6,000 years after the time of closest encounter. This can also be seen in the bottom left panel of Figure~\ref{fig:da_System2_arrow_parabola}. The red, blue, and orange parabola, which represent the distance between TYC 6813-1293-1 and the other three objects in the system lie below the dark grey and dark green parabolas. In other words, at the time of closest encounter TYC 6813-1293-1 is closer to the other three objects than they are to each other. In particular, we think the interaction that causes the disintegration is the one between TYC 6813-1293-1 and TYC 6813-286-1, because the red parabola reaches the minimum and then, shortly thereafter, the disintegration begins (i.e. the dark grey and dark green parabolas start increasing). Only the binary composed of TYC 6813-643-1 A and B survives the encounter, because the separation between them is significantly smaller than the distance between them and TYC 6813-1293-1, so their gravitational bond is strong enough to prevent the disintegration. 

Examining further Figure~\ref{fig:da_System2_arrow_parabola}, we can see that the parabola of objects TYC 6813-643-1 A and B (i.e. the dodger blue parabola) has a downward slope, seemingly implying that the two objects were further away from each other during the close encounter than they are now. We don't think this is real, but we assume it is a result of the fact that the PMs measured by Gaia include the orbital motion, which we cannot remove because we don't know the orbit of the binary. So when we propagate the position back in time this orbital motion causes TYC 6813-643-1 A and B to appear to move closer to each other. We also note that many parabolas reach their minima after the time of closest encounter, and this is probably due to the orbital motion as well. A similar situation we suspect is happening in the left panel of Figure~\ref{fig:da_System2_arrow_parabola} with TYC 6813-286-1. As seen, TYC 6813-286-1 (the blue cross) seems to be outside of the dotted circle at the time of closest encounter, which is the 20,000 AU radius that we used originally to select the candidate disintegrating systems. When we originally selected the candidates we used the TGAS \citep{2016A&A...595A...2G} PMs, but now we are using the Gaia DR3 \citep{2021A&A...649A...1G} PMs. We suspect that because of orbital motion due to the object being an unresolved binary the TGAS \citep{2016A&A...595A...2G} and Gaia DR3 \citep{2021A&A...649A...1G} PMs are slightly different, so the object now seems to be outside of the circle. 

Overall, this is a very interesting system, because it could be a quadruple + binary encounter, leading to the disintegration of the quadruple. 

\subsection{TYC 7240-1438-1, TYC 7240-1159-1 and TYC 7240-850-1}

TYC 7240-1438-1 is an F3V star \citep{1982mcts.book.....H}, TYC 7240-1159-1 is an F8V star \citep{2010PASP..122.1437P}, and TYC 7240-850-1 is a G5V star \citep{2010PASP..122.1437P}. All three objects have approximately solar metallicity (0.0$\pm$0.2, \citealt{2022arXiv220800211G}; 0.05$\pm$0.06, \citealt{2017AJ....153...75K}; 0.07$\pm$0.07, \citealt{2017AJ....153...75K}), although the [Fe/H] measurement for TYC 7240-1438-1 has large uncertainties. The three objects all have different RV from each other and all have good measurements with relatively small uncertainties. TYC 7240-1438-1 has RV = 12.4$\pm$0.3 km s$^{-1}$, TYC 7240-1159-1 has RV = -9.4$\pm$0.5 km s$^{-1}$, and TYC 7240-850-1 has RV = 20.7$\pm$0.2 km s$^{-1}$ \citep[all measurements are from ][]{2022arXiv220800211G}. All three stars have a probability $>$97\% of belonging to the thin disk (see description of methodology above) and, according to BANYAN \citep{2018ApJ...856...23G}, all three stars have a probability of 99.9\% of being field objects. None of the three objects is flagged to be a possible unresolved binary.

\begin{figure*}
    \centering
    \includegraphics[width=\columnwidth]{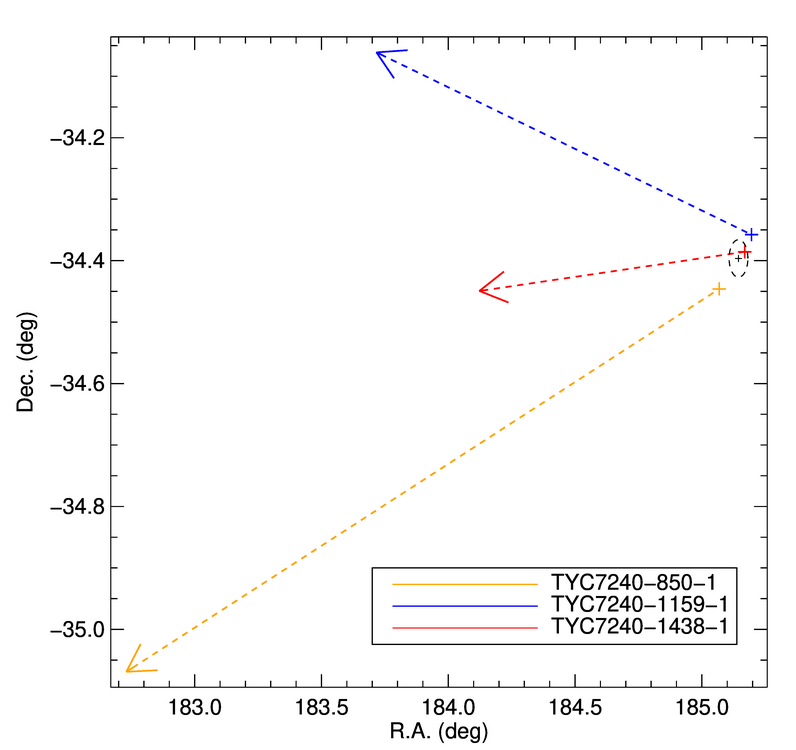}
    \includegraphics[width=\columnwidth]{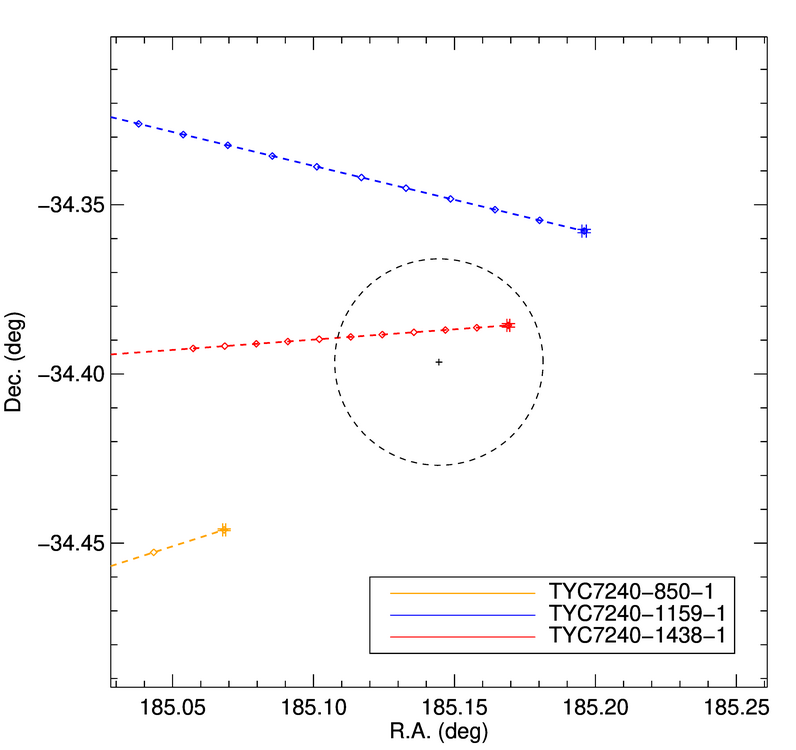}
    \includegraphics[width=\columnwidth]{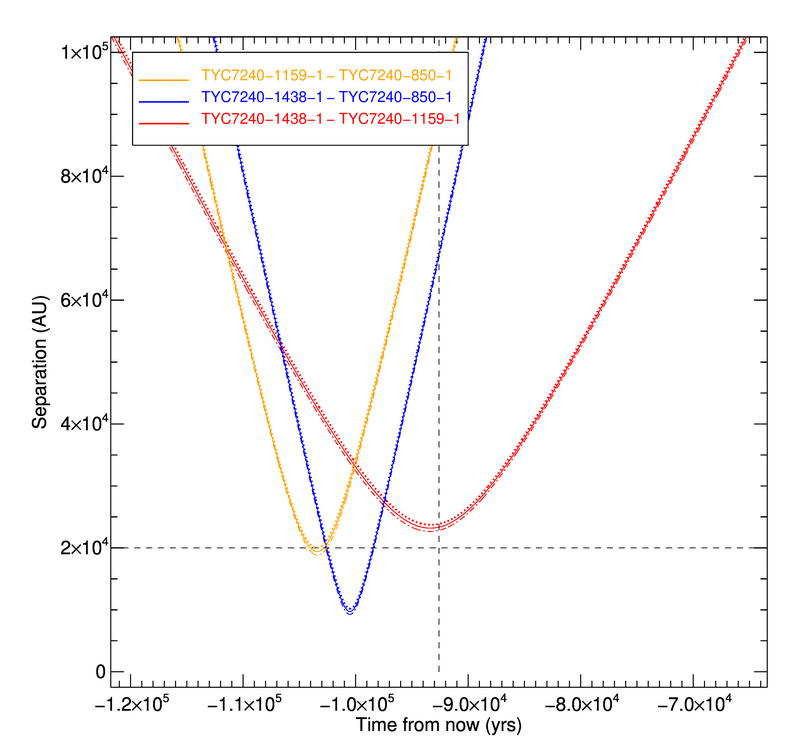}
    \begin{minipage}[b]{\columnwidth}
    \caption{\textit{Top left panel:} Same as the top left panel of Figure~\ref{fig:da_System2_arrow_parabola}, but for the system consisting of TYC 7240-1438-1, TYC 7240-1159-1 and TYC 7240-850-1. TYC 7240-1438-1 passes between the other two objects, so it could be causing their disintegration. \textit{Top right panel:} same as the top right panel of Figure~\ref{fig:da_System2_arrow_parabola} but for the system consisting of TYC 7240-1438-1, TYC 7240-1159-1 and TYC 7240-850-1. All objects in this system have their initial interaction at the time of closest encounter of the system (marked by a cross). As shown in the top left panel, TYC 7240-1438-1 (the red dashed line) passes between TYC 7240-1159-1 (the blue dashed line) and TYC 7240-850-1 (the orange dashed line), possibly triggering the disintegration of the binary. \textit{Bottom left panel:} Same as the bottom left panel of Figure~\ref{fig:da_System2_arrow_parabola}, but for the system consisting of  TYC 7240-1438-1, TYC 7240-1159-1 and TYC 7240-850-1. The separation between TYC 7240-1159-1 and TYC 7240-850-1 (orange parabola) reaches its minimum before the separation between them and TYC 7240-1438-1 (red and blue parabola), meaning that the system was already disintegrating before the close encounter. \label{fig:da_NewSystem_arrow_parabola}}
    \end{minipage}
\end{figure*}

We can see in the top panels of Figure~\ref{fig:da_NewSystem_arrow_parabola} that TYC 7240-1438-1 passes between the other two objects, so it could be the cause of the disintegration of the system. However, we see in the bottom left panel of Figure~\ref{fig:da_NewSystem_arrow_parabola} that the separation between TYC 7240-1159-1 and TYC 7240-850-1 (i.e. the orange parabola) starts increasing before the interaction between TYC 7240-1438-1 and the other two objects. In other words, the orange parabola reaches the minimum before the red and blue parabolas. So, it appears that the possible binary formed by TYC 7240-1159-1 and TYC 7240-850-1 was already disintegrating at the time of closest encounter with TYC 7240-1438-1. This leaves us with two possible explanations -- either TYC 7240-1159-1 and TYC 7240-850-1 were not a binary to begin with but just two objects passing near each other, or TYC 7240-1159-1 and TYC 7240-850-1 were a binary, but their disintegration is caused by another, yet unseen, object.   

\subsection{TYC 4936-84-1 AB, TYC 4933-912-1 AB, and TYC 4934-796-1}

TYC 4936-84-1 is a K1III star \citep{2010PASP..122.1437P} with a slightly low metallicity ([Fe/H] = -0.12$\pm$0.09; \citealt{2017AJ....153...75K}) and a small RV \citep[RV = -1.80$\pm$0.80 km s$^{-1}$$ $;][]{2018A&A...616A...1G}. It has a probability of 99\% of belonging to the thin disk. Gaia DR3 resolves a fainter companion, for which we estimate a spectral type of M4.5V using its absolute $G$ magnitude, parallax, and its colours (i.e. $G$-$BP$, $G$-$RP$, $BP$-$RP$) with Table 5 from \citet{2013ApJS..208....9P}. No metallicity estimate or measurement is available for the companion. We will refer to this binary as TYC 4936-84-1 AB. TYC 4933-912-1 is a K0V star \citep{2010PASP..122.1437P} with a slightly high measured metallicity ([Fe/H] = 0.19$\pm$0.05; \citealt{2021MNRAS.506..150B}) and a measured RV = 15.90$\pm$0.07 km s$^{-1}$ \citep{2021MNRAS.506..150B}. We compute a probability of 98\% of belonging to the thin disk. This object is listed in the SUPERWIDE catalog \citep{2020ApJS..247...66H} as SWB11354. Its companion is a K8V which we will call TYC 4933-912-1 B. The spectral type was estimated by us using the absolute $G$ magnitude, parallax, and the Gaia DR3 colours, as described above. The parallax, PM, and RV of the two objects agree very well \citep[RV = 17.3$\pm$0.8 km s$^{-1}$$ $;][]{2022arXiv220800211G}, so it is highly likely that this object is a real binary, however, the metallicity measurement for TYC 4933-912-1 B is [Fe/H] = 0.02$\pm$0.07 \citep{2017AJ....153...75K}, which is not consistent with the metallicity for TYC 4933-912-1 A. This is puzzling since one would expect that the two components of a binary have the same metallicity since they formed from the same material. Given our analysis in Section 2, we do not think this is due to a systematic difference between GALAH and RAVE leading to a spurious difference between the two components. Another more exotic explanation is that the metallicity of the two objects is different because the two stars did not form together but the K0V captured at some point the K8V. Further followup is needed to clarify the nature of this pair. TYC 4934-796-1 is a K0V \citep{2010PASP..122.1437P} with a high metallicity ([Fe/H] = 0.29$\pm$0.05; \citealt{2021MNRAS.506..150B}), and a measured RV = -19.89$\pm$0.06 km s$^{-1}$ \citep{2021MNRAS.506..150B}. Gaia DR2 reports a RV of $ -19.2\pm0.5$ km s$^{-1}$, while \citet{2022A&A...659A..95T} reports a RV of $ -20.8\pm0.7$ km s$^{-1}$. Both measurements are consistent with the one from \citet{2021MNRAS.506..150B}. It has a probability of 99\% of belonging to the thin disk, regardless of which RV we use for the calculation. All of these five objects have a probability of 99.9\% of being field objects according to BANYAN \citep{2018ApJ...856...23G}. 

\begin{figure*}
    \centering
    \includegraphics[width=\columnwidth]{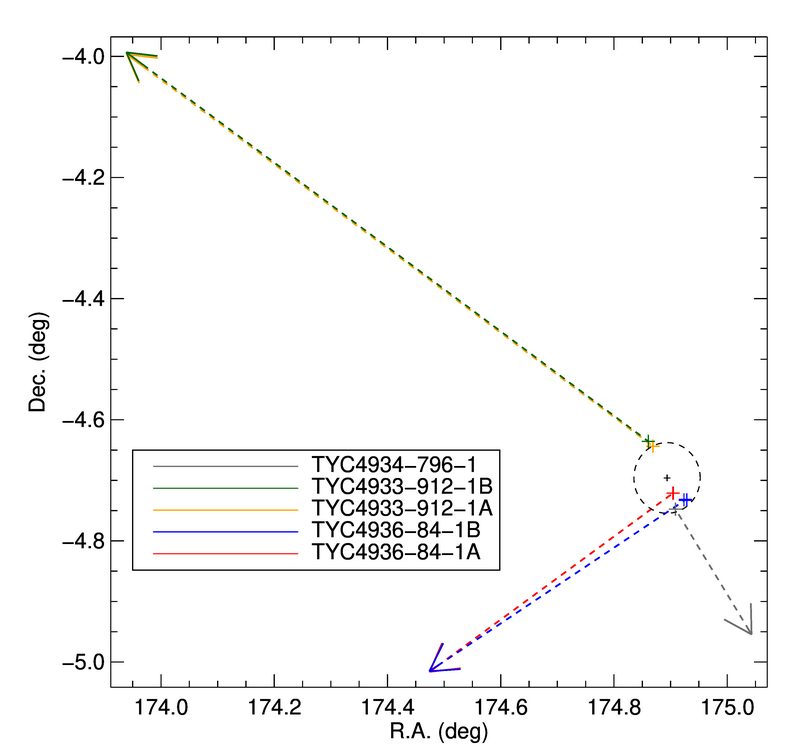}
    \includegraphics[width=\columnwidth]{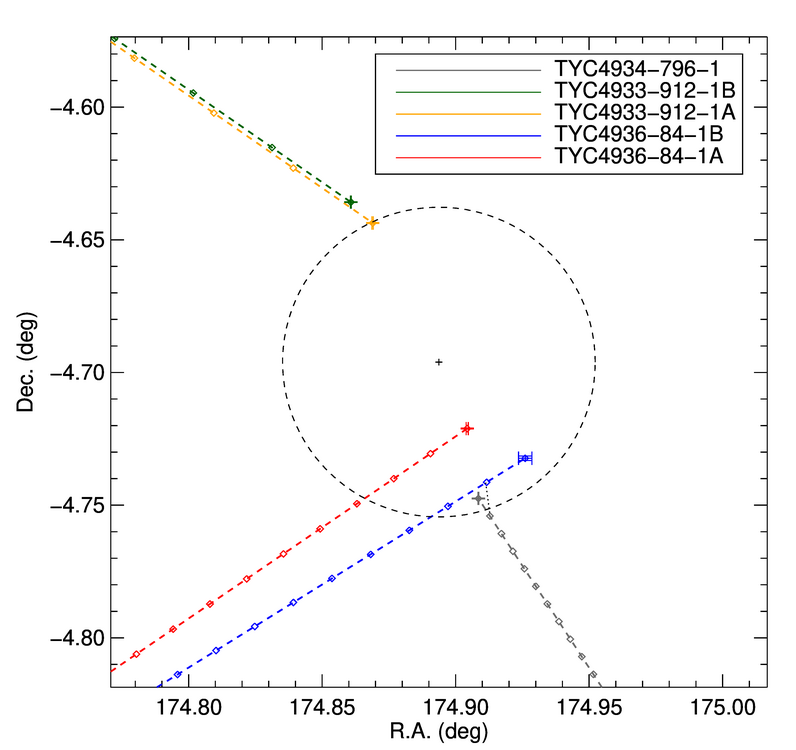}
    \includegraphics[width=\columnwidth]{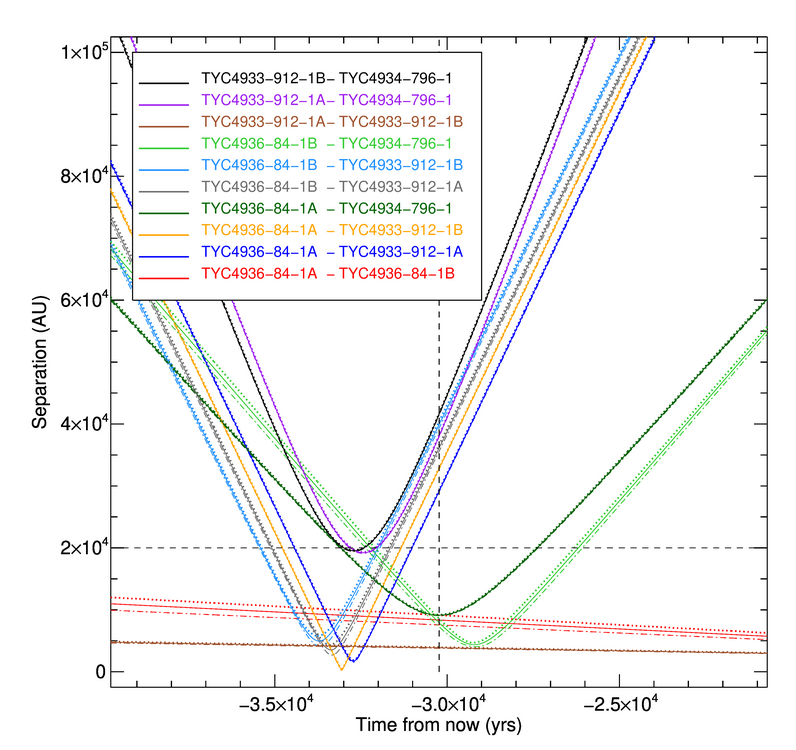}
    \begin{minipage}[b]{\columnwidth}
    \caption{\textit{Top left panel:} Same as the top left panel of Figure~\ref{fig:da_System2_arrow_parabola}, but for the system consisting of TYC 4936-84-1 AB, TYC 4933-912-1 AB, and TYC 4934-796-1. The direction of motion of the two binaries and TYC 4934-796-1 are almost perpendicular to each other, so it is likely that these systems were unrelated to each other, and this is just a close encounter without any visible disintegration. \textit{Top right panel:} same as the top right panel of Figure~\ref{fig:da_System2_arrow_parabola}, but for the system consisting of TYC 4936-84-1AB, TYC 4933-912-1AB and TYC 4934-761-1. All objects in this system have their initial interaction at the time of closest encounter of the system (marked by a cross) except for TYC 4936-84-1B (the blue dashed line) and TYC 4934-796-1 (the dark grey dashed line) which first interacted around 1,000 years after the system closest encounter. This candidate system appears to be just a chance alignment. \textit{Bottom left panel:} Same as the bottom left panel of Figure~\ref{fig:da_System2_arrow_parabola}, but for the system consisting of TYC 4936-84-1 AB, TYC 4933-912-1 AB, and TYC 4934-796-1. The two binaries have a close encounter first (i.e. the blue, orange, dark grey, and dodger blue parabola reach their minima first), then the separation between TYC 4933-912-1 AB and TYC 4934-796-1 start increasing (purple and black parabola), so this could be a sign of the disintegration of the system. However, the current direction of motion of TYC 4933-912-1 AB and TYC 4934-796-1 are almost opposite to each other, and it is unlikely for an interaction to have caused such a dramatic change of path. TYC 4936-84-1 AB and TYC 4933-912-AB remain bound as binaries, however their parabolas (red and brown) appear to show that the components are moving closer to each other. We attribute this to unaccounted orbital motion (see Section 5.3 for further discussion)} \label{fig:da_System4_arrow_parabola}
    \end{minipage}
\end{figure*}

The true nature of this system is unclear. Looking at the top panels of Figure~\ref{fig:da_System4_arrow_parabola} we can see that the two binaries (TYC 4936-84-1 AB and TYC 4933-912-1 AB) and TYC 4934-796-1 are going in three different directions, almost perpendicular to each other. This can be interpreted as a simple close encounter between three unrelated systems, with no visible disintegration occurring. However, it is still possible that there could be unseen low-mass objects ejected by one of the three systems. Another possibility is that TYC 4933-912-1 AB and TYC 4934-796-1 used to be a triple system, and now, after the interaction with TYC 4936-84-1 AB, they are disintegrating. This hypothesis comes from the fact that TYC 4934-796-1 and TYC 4933-912-1 A have metallicity which is consistent with each other within their errors but, as we discussed above, the metallicity of TYC 4933-912-1 A and B need further investigation to explain their inconsistency. The bottom left panel of Figure~\ref{fig:da_System4_arrow_parabola} could help explain the nature of this system. The red and brown parabolas represent TYC 4936-84-1 AB and TYC 4933-912-1 AB, which are still bound together, so their separation does not increase with time. We note that both parabolas have a shallow slope to them, most likely due to the unaccounted orbital motion of the pairs. In practice, since the Gaia astrometric solution assumes that these objects are singles and not binaries, it does not separate the PM of the object from its orbital motion. The measured PM in the Gaia catalogue is, therefore, the sum of the PM and the orbital motion, and this makes it appear as if the two objects are moving closer to each other. Figure~\ref{fig:da_System4_arrow_parabola} shows that the two binaries interact first, i.e. the blue, orange, dark grey, and dodger blue parabola reach their minima first, with minimum separation of $\lesssim$5,000 AU. This could be causing the ejection of TYC 4934-796-1 from the TYC 4933-912-1 AB system. The separation between these three objects, represented by the purple and black parabola, starts increasing just after the interaction between the two binaries. However, the current direction of motion for TYC 4933-912-1 AB and TYC 4934-796-1 are almost exactly opposite to each other, and it is unlikely for the interaction to have caused such a large change of course for an object as massive as a K0V star. Finally, TYC 4934-796-1 has a close encounter with the TYC 4936-84-1 AB (i.e. the dark green and lime green parabola) approximately 3,000 years after the disintegration began or approximately 1,000 years after the time of closest encounter of the full system (see bottom left and top right panels of Figure~\ref{fig:da_System4_arrow_parabola}). 

Further analysis of this system, in particular of the discrepancy between the metallicity of TYC 4933-912-1 A and B could help clarify the nature of this system.

\subsection{TYC 9281-3037-1, TYC 9281-2422-1, and TYC 9281-1175-1AB?}

TYC 9281-3037-1 is a G0V star \citep{2010PASP..122.1437P} with a very low metallicity of -2.3$\pm$0.1 \citep{2021MNRAS.506..150B} and a very high PM and RV \citep[238.1$\pm$0.9 km s$^{-1}$$ $,][]{2021MNRAS.506..150B}. Because of its high PM and RV, this object has a probability of 100\% of belonging to the halo (see methodology above), which is consistent with its low metallicity. We found that Gaia DR3 has a slightly discrepant RV = 230.4$\pm$2.4 km s$^{-1}$, but even if we assume the Gaia value instead of the \citet{2021MNRAS.506..150B} value the probability to belong to the halo remains 100\%. TYC 9281-2422-1 is a G5V star \citep{2010PASP..122.1437P} with an estimated metallicity which is consistent with solar, but with large uncertainties ([Fe/H] = 0.2$\pm$0.4; \citealt{2006ApJ...638.1004A}). It has a relatively low RV, but its measurement also has large uncertainty \citep[RV = 16$\pm$6 km s$^{-1}$$ $,][]{2022arXiv220800211G}. We speculate that this large uncertainty could be due to this object being an unresolved binary, i.e. the large uncertainty could be a result of line broadening due to the orbital velocity of the pair with the orbital plane being almost aligned with the line of sight. Gaia does not resolve the possible system. Despite the large RV uncertainty, this object has a probability of 99\% of belonging to the thin disk. TYC 9281-1175-1A is an F8V star \citep{2010PASP..122.1437P} with a well measured metallicity of 0.06$\pm$0.08 \citep{2021MNRAS.506..150B} and a well measured RV = -5.2$\pm$0.1 km s$^{-1}$ \citep{2021MNRAS.506..150B}. Gaia DR3 reports a RV of $ -5.0\pm0.2$ km s$^{-1}$ and \citet{2022A&A...659A..95T} reports a RV of $ -4.9\pm0.3$ km s$^{-1}$, which are consistent with the \citet{2021MNRAS.506..150B} value. Gaia resolves the object in two separate sources, but does not provide parallax nor PM measurement for the secondary, so we cannot conclude if this secondary source is a real companion or just a chance alignment with a background source. If we assume that the secondary is at the same distance as the primary, then we can use its Gaia $G$ magnitude to estimate its spectral type to be M5V. TYC 9281-1175-1AB?  has a probability of 99\% of belonging to the thin disk, regardless of which RV we use for the calculation. According to BANYAN \citep{2018ApJ...856...23G} all of the stars are members of the field with a probability of 99.9\%. 

\begin{figure*}
    \centering
    \includegraphics[width=\columnwidth]{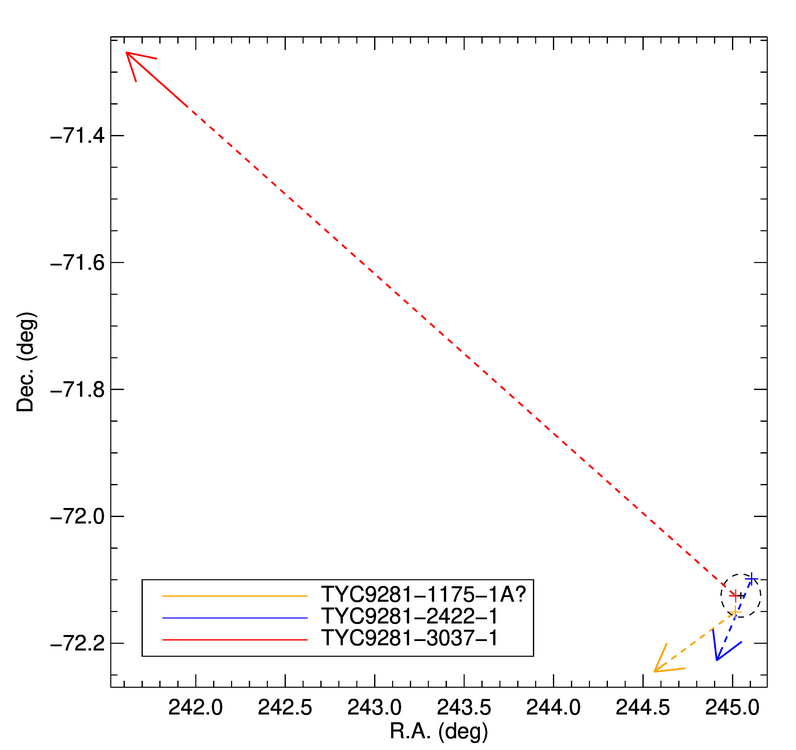}
    \includegraphics[width=\columnwidth]{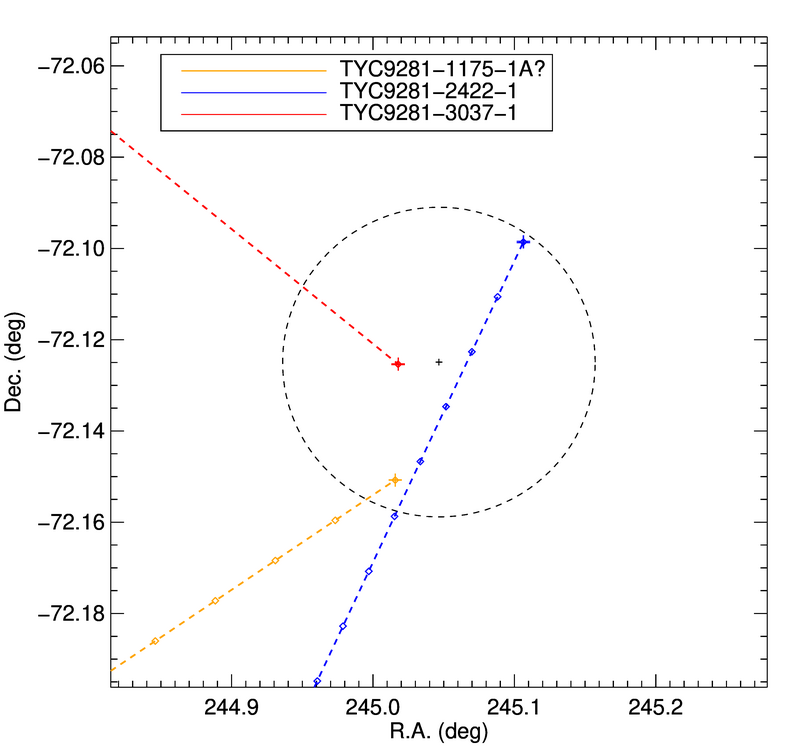}
    \includegraphics[width=\columnwidth]{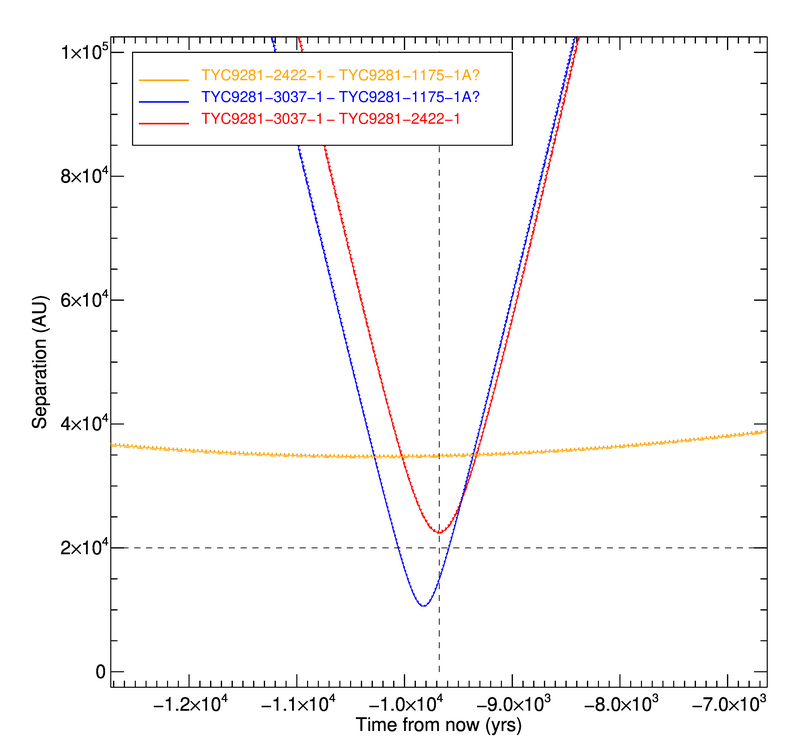}
    \begin{minipage}[b]{\columnwidth}
    \caption{\textit{Top left panel:} Same as the top left panel of Figure~\ref{fig:da_System2_arrow_parabola}, but for the system consisting of TYC 9281-3037-1, TYC 9281-2422-1, and TYC 9281-1175-1A?. The possible unresolved companion to TYC 9281-1175-1A? is not shown because there is no PM measurement available for it. TYC 9281-3037-1 passes between the other two objects, so it could be causing their disintegration. \textit{Top right panel:} same as the top right panel of Figure~\ref{fig:da_System2_arrow_parabola}, but for the system consisting of TYC 9281-3037-1, TYC 9281-2422-1 and TYC 9281-1175-1A?. All objects in this system have their initial interaction at the time of closest encounter of the system (marked by a cross). The fast moving TYC 9281-3037-1 leaves the area shown in less than 1,000 yr, so there is no red diamond in the plot. As shown in the top left panel, TYC 9281-3037-1 (the red dashed line) passes between TYC 9281-2422-1 (the blue dashed line) and TYC 9281-1175-1A? (the orange dashed line), so it is likely to be the cause of the disintegration of the binary/triple. \textit{Bottom left panel:} Same as the bottom left panel of Figure~\ref{fig:da_System2_arrow_parabola}, but for the system consisting of TYC 9281-3037-1, TYC 9281-2422-1, and TYC 9281-1175-1A?. The possible unresolved companion to TYC 9281-1175-1A? is not shown because there is no PM measurement available for it. TYC 9281-3037-1 passes between the other two objects, as shown by the red and blue parabolas reaching a smaller separation than the orange parabola. This interaction could be the cause of the disintegration of the system. The disintegration in this case is ``gentle'' as shown by the orange parabola being almost flat, i.e. the separation between TYC 9281-2422-1 and TYC 9281-1175-1A? increases slowly. \label{fig:da_SystemX_arrow_parabola}}
    \end{minipage}
\end{figure*}

We speculate that TYC 9281-2422-1 and TYC 9281-1175-1AB? used to form a binary because their similar [Fe/H] and somewhat similar kinematic, and that TYC 9281-3037-1, a halo star, caused their breakup. Looking at the top panels of Figure~\ref{fig:da_SystemX_arrow_parabola} we can see that TYC 9281-3037-1 (red arrow and dashed line) passes between TYC 9281-2422-1 (blue arrow and dashed line) and TYC 9281-1175-1A (orange arrow and dashed line), so this interaction could be the cause of the disintegration. The bottom left panel of the same Figure confirms this, as we can see that TYC 9281-3037-1 comes quite close with TYC 9281-1175-1A first (i.e. the blue parabola reaches its minimum first), then interacts with TYC 9281-2422-1 (i.e. the red parabola reaches its minimum second). We can also see that this appears to be a more ``gentle'' breakup compared to the other systems studied here, since the separation between TYC 9281-2422-1 and TYC 9281-1175-1A increases slowly (i.e. the orange parabola is almost flat). 

Deep AO imaging and/or RV monitoring of TYC 9281-2422-1 and TYC 9281-1175-1AB? could be useful to understand if either of those object is itself a tight binary.

\subsection{TYC 7731-1951-1, TYC 7731-2128-1AB, and TYC 7731-1995-1ABC?.}

TYC 7731-1951-1 is a G5V star \citep{2010PASP..122.1437P} with a low estimated metallicity, but with large uncertainties ([Fe/H] = -0.3$\pm$0.2; \citealt{2022arXiv220800211G}). Using its well measured RV \citep[RV = 32.6$\pm$0.3 km s$^{-1}$$ $;][]{2022arXiv220800211G} we find that this object has a probability of 97\% of belonging to the thin disk. TYC 7731-2128-1 is a G0V star \citep{2010PASP..122.1437P} with an estimated metallicity of [Fe/H] = 0.3$\pm$0.3 \citep{2006ApJ...638.1004A}. This star has a low RV = 0.6$\pm$0.2 km s$^{-1}$ \citep{2022arXiv220800211G} but because of its high PM, it has a probability of 76\% of belonging to the thick disk, and only 24\% of belonging to the thin disk. We found that it has a close companion resolved by Gaia \citep{2021A&A...649A...1G}, with parallax and PM consistent with being a companion to TYC 7731-2128-1. So we call the primary TYC 7731-2128-1 A and this new companion TYC 7731-2128-1 B. We estimate the spectral type of the new companion using its absolute $G$ magnitude, parallax, and the Table 5 from \citet{2013ApJS..208....9P}, and find that this is an M3V. There are no [Fe/H] nor RV measurements for TYC 7731-2128-1 B. TYC 7731-1995-1A is a K0III \citep{1978mcts.book.....H}. The estimated [Fe/H] is 0.2$\pm$0.2 \citep{2022arXiv220800211G}, and its RV is 5.3$\pm$0.1 km s$^{-1}$ \citep{2018A&A...616A...1G}. Using the parallax, PM, and RV we estimate a probability of 99\% of belonging to the thin disk. This object is also in the catalog of accelerations \citep{2021ApJS..254...42B} indicating that it could be an unresolved binary. Its measured RV has small uncertainties, but this could be due to the fact that the orbital plane of the binary is perpendicular to the line of sight, so the orbital velocity does not contribute much to the RV uncertainty. Using ALADIN \citep{2000A&AS..143...33B}, we see that there is another object very close to this target, but Gaia DR3 \citep{2021A&A...649A...1G} does not report the parallax nor the PM of this other object, so we cannot establish if this new object is a companion to TYC 7731-1995-1A (hence the cause of its acceleration) or just a background star. If we assume that this new star is at the same distance as the K0III, we estimate its spectral type to be M3V. According to BANYAN, TYC 7731-1951-1, TYC 7731-2128 AB, and TYC 7731-1995-1ABC? are all field objects.

\begin{figure*}
    \centering
    \includegraphics[width=\columnwidth]{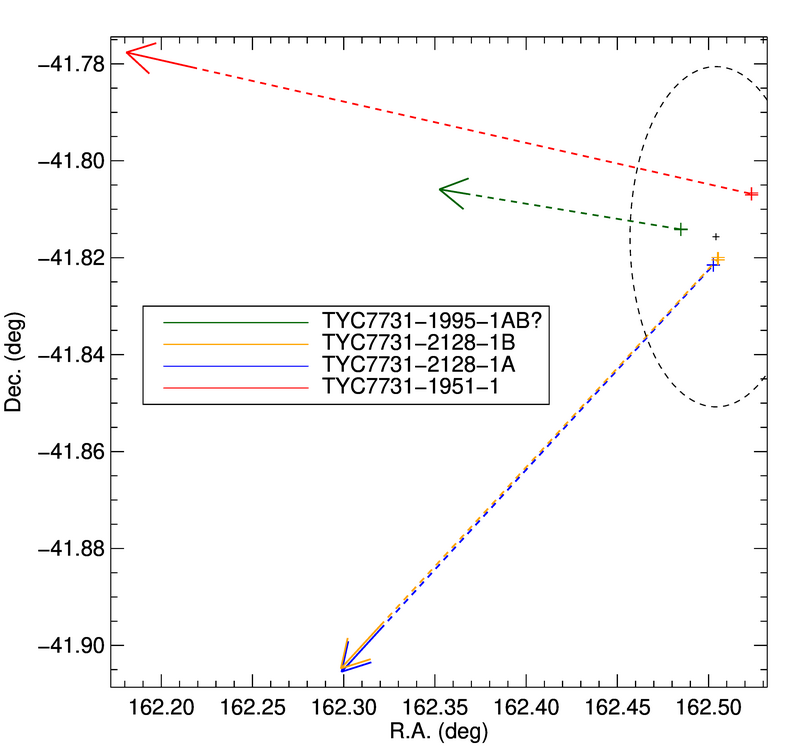}
    \includegraphics[width=\columnwidth]{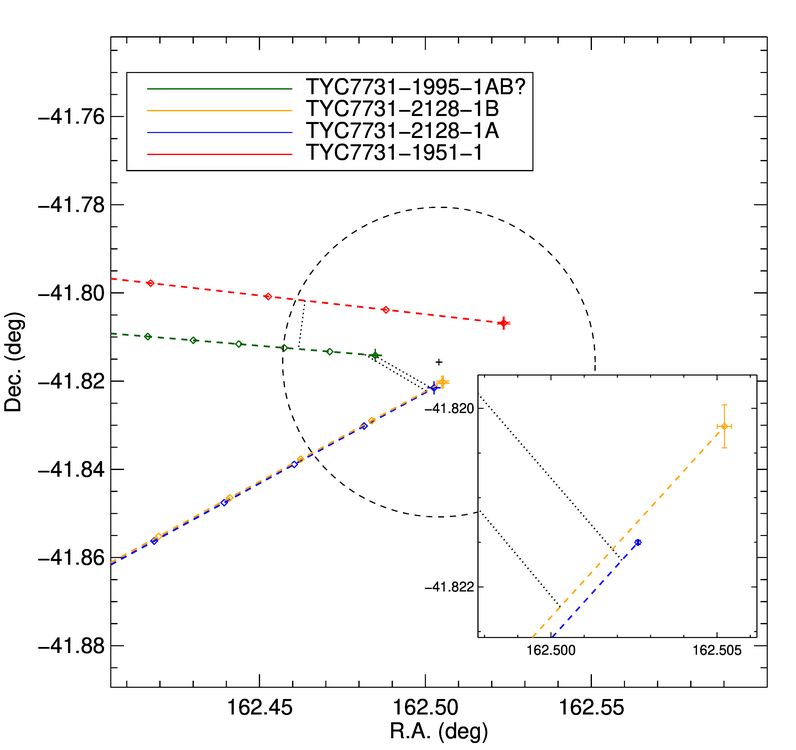}
    \includegraphics[width=\columnwidth]{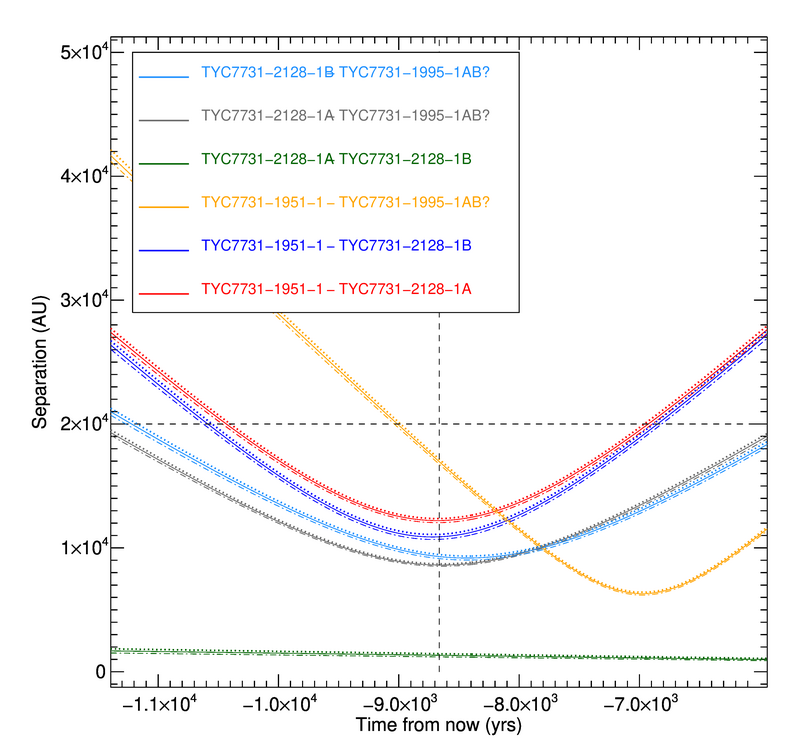}
    \begin{minipage}[b]{\columnwidth}
        \caption{\textit{Top left panel:} Same as the top left panel of Figure~\ref{fig:da_System2_arrow_parabola}, but for the system consisting of TYC 7731-1951-1, TYC 7731-2128 AB, and TYC 7731-1995-1AB?. The possible companion to TYC 7731-1995-1AB? is not shown because there is no PM measurement available for it. The interaction between TYC 7731-1951-1 and TYC 7731-2128 AB is likely to cause the breakup of the TYC 7731-2128 AB and TYC 7731-1995-1AB? system. \textit{Top right panel:} same as the top right panel of Figure~\ref{fig:da_System2_arrow_parabola}, but for the system consisting of TYC 7731-1951-1, TYC 7731-2128-1AB and TYC 7731-1995-1AB?. All objects in this system have their initial interaction at the time of closest encounter of the system (marked by a cross) except for TYC 7731-1951-1 (the red dashed line) and TYC 7731-1995-1AB? (the dark green dashed line) which interact just before the 2,000 years mark, TYC 7731-2128-1A (the blue dashed line) and TYC 7731-1995-1AB? as well as TYC 7731-2128-1B (the orange dashed line) and TYC 7731-1995-1AB? which have their interaction shortly after the closest encounter of the system. TYC 7731-1951-1 (red dashed line) is likely causing the disintegration of TYC 7731-2128-1AB and TYC 7731-1995-1AB? (blue, orange and dark green dashed lines) by first interacting with TYC 7731-2128-1AB, then later interacting with TYC 7731-1995-1AB? (the dark green dashed line). \textit{Bottom left panel:} Same as the bottom left panel of Figure~\ref{fig:da_System2_arrow_parabola}, but for the system consisting of TYC 7731-1951-1, TYC 7731-2128 AB, and TYC 7731-1995-1AB?. The possible companion to TYC 7731-1995-1AB? is not shown because there is no PM measurement available for it. TYC 7731-1951-1 passes very close to TYC 7731-2128 AB (red and blue parabolas) likely triggering the disintegration of the TYC 7731-2128 AB and TYC 7731-1995-1AB? system (dark grey and dodger blue parabolas). TYC 7731-2128 AB survives the encounter (dark green parabola). \label{fig:da_System8_arrow_parabola}}
    \end{minipage}
\end{figure*}

Even though we do not have good [Fe/H] measurement to establish which object were previously associated, we can still use Figure~\ref{fig:da_System8_arrow_parabola} to interpret the close encounter. The top panels of Figure~\ref{fig:da_System8_arrow_parabola} show that TYC 7731-1951-1 comes close to TYC 7731-2128 AB and TYC 7731-1995-1AB? and then move on. The bottom left panel of Figure~\ref{fig:da_System8_arrow_parabola} gives us a more complete picture of the interaction. TYC 7731-1951-1 interacts with TYC 7731-2128 AB (red and blue parabolas) but comes closer to TYC 7731-2128 B. Shortly thereafter, the disintegration between TYC 7731-2128 AB and TYC 7731-1995-1AB? begins (dark grey and dodger blue parabolas, see also the inset of the top right panel). So we conclude that the interaction between TYC 7731-1951-1 and TYC 7731-2128 AB is the cause of the disintegration of the triple system (or maybe quadruple). We note that the dark grey and dodger blue parabola are slightly different from each other even though in theory they should be identical (because TYC 7731-2128 A and B are binary) and we think that this difference is due to the orbital motion of the binary. We can also see that TYC 7731-2128 AB is still a binary because the separation between the two components is constant (dark green parabola). We can also see that TYC 7731-1951-1 and TYC 7731-1995-1AB? come close to each other $\sim$2000 yr after the disintegration began, so this interaction cannot be the cause (see the orange parabola in the bottom left panel as well as the top right panel of Figure~\ref{fig:da_System8_arrow_parabola}). 

Better metallicity measurements for all objects in this system would help us confirm the scenario we think is likely. A measurement of the parallax and PM for the newly discovered object near TYC 7731-1995-1AB? is needed to find out is these two objects form a binary or if it is just a chance alignment.

\begin{figure*}
    \centering
    \includegraphics[width=\columnwidth]{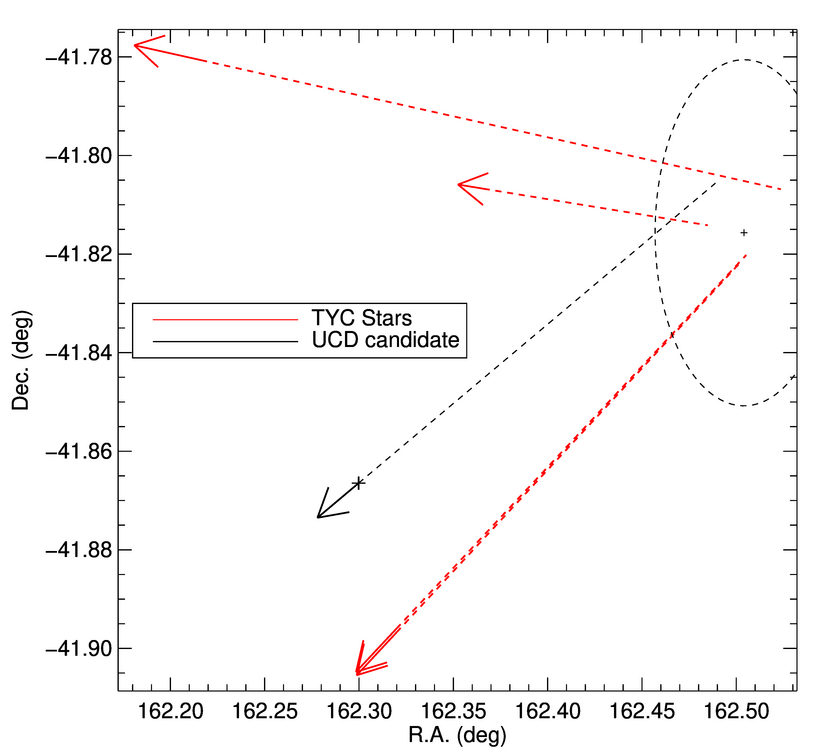}
    \includegraphics[width=\columnwidth]{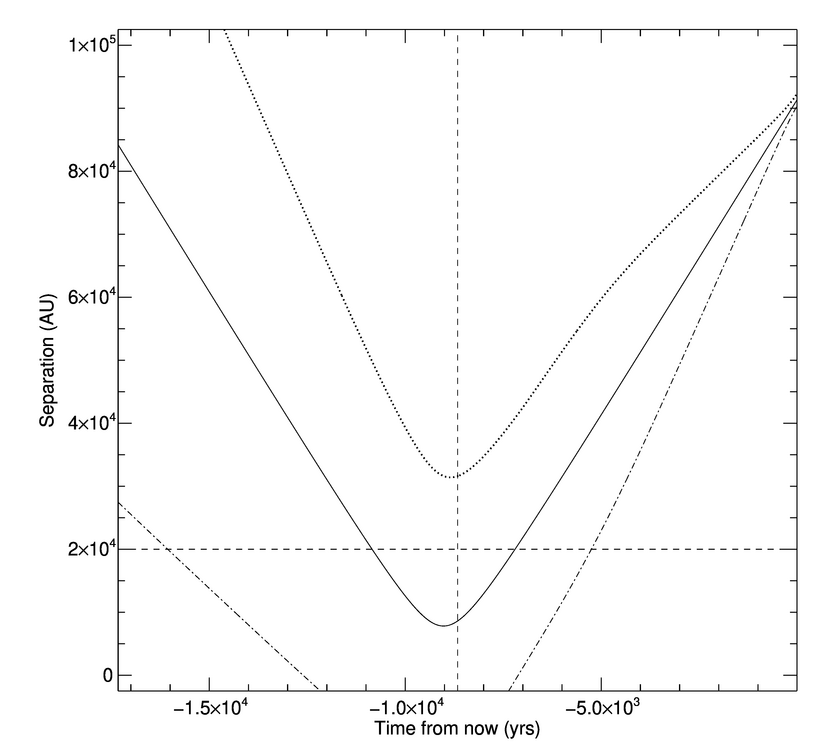}
    \caption{\textit{Left panel:} Same as the top left panel of Figure~\ref{fig:da_System8_arrow_parabola}, but highlighting the possible UCD ejected component discovered here. The TYC stars are plotted in red, while the UCD is plotted in black. \textit{Right panel:} Same as the bottom left panel of Figure~\ref{fig:da_System8_arrow_parabola}, but showing only the possible new UCD ejected component discovered here. The solid parabola shows the separation as a function of time between the UCD and the center point of the disintegrating system at the time of closest encounter. The dotted and dash-dot parabolas are the upper and lower 1$\sigma$ uncertainties.}
    \label{fig:da_System8_object85}
\end{figure*}

The analysis described in Section 4 revealed a possible additional ejected UCD component of this system. The left panel of Figure~\ref{fig:da_System8_object85} shows that the PM for the new UCD is well aligned with the PM of TYC 7731-2128-1 AB, suggesting that the UCD candidate could have formed a triple system with those two stars. We note that the PM measurement for this UCD is based on the DECam observation combined with the VHS J data. The right panel of Figure~\ref{fig:da_System8_object85} shows that the parabola for the UCD reaches its minimum very close to the time of closest encounter, marked by a vertical dashed line. Even though the errors on the UCD candidate parabola are larger than those on the TYC stars, the maximum separation at the time of closest encounter is $\lesssim3.5\times10^4$\,AU, therefore it is plausible that this object was initially part of the system and is being ejected as a result of the disintegration of the system. The UCD is well detected in our DECam $z$ image, but only barely detected in the $Y$ image. Moreover, the UCD candidate is also well detected in the VHS $J$ and $K$ data, as well as the WISE $W1$ data, but only barely detected in the $W2$ image. It is undetected in Gaia DR3. Further inspection of DSS2 $B$ images reveal a faint detection at the location of the UCD. This detection is puzzling and in disagreement with the fact that this source does not appear in Gaia DR3, since the Gaia $G$ and $G_{\rm BP}$ filters both overlap the DSS2 $B$ filter, and Gaia DR3 is much deeper than DSS2. We propose three possible explanations for this puzzling detection. The first possible explanation is that the DSS2 detection is spurious, i.e. an image artifact. The second possible explanation is that the DSS2 detection is real, but it is not associated with our UCD candidate. The fact that this source does not appear in Gaia DR3 can be explained assuming that the DSS2 source is a transient (e.g. a nova, or a long-period variable source). If this is the case, since this source is only detected in DSS2 $B$ but not in DSS2 $R$ and $IR$, we can assume that it does not contaminate the photometry and astrometry of our UCD, which is based only on near-infrared data. The third possible explanation is that the DSS2 detection is real and it is actually associated with our UCD candidate. This would mean that our candidate is not a UCD but instead some background object with peculiar colours, and probably variable at short wavelength. Using the available photometry (excluding the puzzling DSS2 $B$ detection), assuming that the object is at the same distance as the TYC stars, and employing the method described in Section 4, we find that this UCD candidate is likely an early-L dwarf. Looking at the top panel of Figure~\ref{fig:system8_object85_CMD} we can see that given the relatively large error bars the $z-Y$ colour is consistent with spectral type M7--L4, while the $Y-J_{\rm VHS}$ colour is consistent with spectral type M6--L1 (but we note that the 1-$\sigma$ error bars for the UCD candidate also overlap the 1-$\sigma$ scatter range for M5 and L2). The bottom panel of Figure~\ref{fig:system8_object85_CMD} shows that the $(J-K)_{\rm MKO}$ colour is consistent with spectral type M8--L4. The strongest constraint on the spectral type comes from the $M_{J\rm MKO}$ which is consistent with L3. This estimate however is a bit inconsistent with the estimate that comes from the $Y-J_{\rm VHS}$ colour. Overall, we conclude that this could be a slightly underluminous and slightly blue L2--L3 dwarf. Its properties are listed in Table~\ref{tab:system8_newLdwarf}. Additional imaging would help to further improve the photometry and the PM measurement for this object, and spectroscopy is desirable to confirm or refute its nature, thus clarifying the mystery surrounding the DSS2 $B$ detection.

If this additional object is confirmed to be a UCD, it would be the first example of an ejected UCD benchmark.

\begin{table}
    \centering
    \begin{tabular}{c c c c}
    Property & Value & Units & Ref. \\
    \hline
    R.A. & 162.299799$\pm(12\times10^{-5})$ & (deg) & This paper \\
    Dec. & -41.866472$\pm(3.4\times10^{-5})$ & (deg) & This paper \\
    $\mu_\alpha^*$ & $-59\pm24$ & (mas yr$^{-1}$) & This paper \\
    $\mu_\delta$   & $-25\pm24$ & (mas yr$^{-1}$) & This paper \\
    $z$ & 21.93$\pm$0.13 & (mag) & This paper \\
    $Y$ & 21.08$\pm$0.24 & (mag) & This paper \\
    $J$ & 19.42$\pm$0.16 & (mag) & VHS \\
    $K$ & 18.08$\pm$0.23 & (mag) & VHS \\
    $W1$ & 16.862$\pm$0.043 & (mag) & CatWISE2020 \\
    $W2$ & 17.37$\pm$0.21 & (mag) & CatWISE2020 \\
    \hline
    \end{tabular}
    \caption{Coordinates, PM, and photometry for the new candidate UCD.}
    \label{tab:system8_newLdwarf}
\end{table}

\section{Conclusion and Future work}

We applied the method first described in \citet{2016csss.confE.137Y} to the TGAS catalogue \citep{2015A&A...574A.115M} to search for candidate disintegrating systems. In particular, in this paper we have focused our attention onto 5 candidate disintegrating systems, further characterizing their nature and searching for additional low-mass components that were missed by Gaia. Three of the disintegrating systems presented here are in our opinion very promising (see section 5.1, 5.4 and 5.5), in particular the group consisting of TYC 7731-1951-1, TYC 7731-2128 AB, and TYC 7731-1995-1ABC? since it includes an additional UCD candidate. The other two candidate systems are slightly more puzzling. This is because the candidate system consisting of TYC 7240-1438-1, TYC 7240-1159-1 and TYC 7240-850-1 have no clear indication of which object causes the disintegration. The candidate system consisting of TYC 4936-84-1 AB, TYC 4933-912-1 AB, and TYC 4934-796-1 have significant discrepancy in the metallicity of its components, and the PM and separation plots do not show a well defined interaction scenario compared to the other systems. All 5 disintegrating systems listed in this paper would require follow-up observations to obtain high-quality parallax, PM, RV and [Fe/H] measurements for all members. For objects in the Catalog of Acceleration, adaptive optics observations or RV monitoring are desirable to verify if they are real binaries or multiple systems. Regarding the candidate system consisting of TYC 7731-1951-1, TYC 7731-2128 AB, and TYC 7731-1995-1ABC?, spectroscopy is needed to confirm the nature of the candidate UCD. Subsequently, we would obtain a good measurement of the parallax to validate that the UCD is part of this candidate disintegrating system. Furthermore, to substantiate that these candidate systems are indeed disintegrating due to their close encounter, dynamical simulations to examine their interaction is also essential. In addition, now that Gaia DR3 \citep{2022arXiv220800211G} is publicly accessible we will continue the expansion of our list to study with even higher quality candidate systems. We would also then select photometric candidates from PanSTARRS \citep{2016arXiv161205560C}, SDSS \citep{2000AJ....120.1579Y}, UKIDSS \citep{2007MNRAS.379.1599L}, UHS \citep{2018MNRAS.473.5113D}, VHS \citep{2021yCat.2367....0M} and WISE \citep{2010AJ....140.1868W} for possible UCD companion/s to this list of disintegrating systems. The continuation of this work should help us to further fine tune our method to be able to identify all components during disintegration. This then would allow us to place new constraints onto the typically invisible planetary mass objects within the multiple systems. Moreover, determining the observed frequency of disintegrators will allow us to uncover the rate at which this type of systems interact in the Galactic disk. The fundamental importance of searching for possible disintegrating multiple systems is its contribution in providing further constraints to the formation models of binaries and multiple systems. Besides that, it has always been one of the main research interests among the low-mass community to understand the dominant formation mechanism of ultra-cool objects. Observational constraints are vital, especially on the binary fraction and the initial mass function, in order to test the various formation models that have been proposed \citep[e.g.][]{2012MNRAS.427.1182S}. The answer to perhaps the initial requirements for capture and the origin of some of the free-floating planets could also be provided from this work.

\section*{Acknowledgements}
We thank the anonymous referee for their thorough review and for their helpful comments that improved the quality of this manuscript. AKPY would like to thank the Proyecto CONICYT-REDES140024, ``SOCHIAS grant through ALMA/Conicyt Project \#31150039'' and Ministry of Economy, Development, and Tourism's Millennium Science Initiative through grant IC120009, awarded to The Millennium Institute of Astrophysics, MAS for financial support. MG is supported by the EU Horizon 2020 research and innovation programme under grant agreement No 101004719. JAC-B acknowledges support from FONDECYT Regular N 1220083. This work presents results from the European Space Agency (ESA) space mission Gaia. Gaia data are being processed by the Gaia Data Processing and Analysis Consortium (DPAC). Funding for the DPAC is provided by national institutions, in particular the institutions participating in the Gaia MultiLateral Agreement (MLA). The Gaia mission website is \url{https://www.cosmos.esa.int/gaia}. The Gaia archive website is \url{https://archives.esac.esa.int/gaia}. This project used data obtained with the Dark Energy Camera (DECam), which was constructed by the Dark Energy Survey (DES) collaboration. Funding for the SDSS and SDSS-II has been provided by the Alfred P. Sloan Foundation, the Participating Institutions, the National Science Foundation, the U.S. Department of Energy, the National Aeronautics and Space Administration, the Japanese Monbukagakusho, the Max Planck Society, and the Higher Education Funding Council for England. The SDSS Web Site is \url{http://www.sdss.org/}. This publication makes use of data products from the Wide-field Infrared Survey Explorer, which is a joint project of the University of California, Los Angeles, and the Jet Propulsion Laboratory/California Institute of Technology, funded by the National Aeronautics and Space Administration. This research has made use of the SIMBAD database, operated at CDS, Strasbourg, France. This work is based in part on data obtained as part of the UKIRT Infrared Deep Sky Survey.Based on observations obtained as part of the VISTA Hemisphere Survey, ESO Progam, 179.A-2010 (PI: McMahon).

\section*{Data Availability}
The DECam observations are publicly available via the NOIRLab archive (\url{https://astroarchive.noirlab.edu/}). All survey data used in this paper is publicly available through the relevant survey archives. 




\bibliographystyle{mnras}
\bibliography{Paper_1_astroph} 








\bsp	
\label{lastpage}
\end{document}